\documentclass[fleqn,usenatbib]{mnras}

\usepackage{newtxtext,newtxmath}

\usepackage[T1]{fontenc}

\DeclareRobustCommand{\VAN}[3]{#2}
\let\VANthebibliography\thebibliography
\def\thebibliography{\DeclareRobustCommand{\VAN}[3]{##3}\VANthebibliography}

\usepackage{graphicx}	
\usepackage{amsmath}	
\usepackage{tikz}

\usepackage[T1]{fontenc}
\usepackage{aecompl}
\usepackage{calligra}
\DeclareMathAlphabet{\mathcalligra}{T1}{calligra}{m}{n}
\DeclareFontShape{T1}{calligra}{m}{n}{<->s*[2.2]callig15}{}

\usepackage{hyperref}
\hypersetup{
    pdfnewwindow=true,      
    colorlinks=true,       
    linkcolor=orange,      
    citecolor=cyan,        
    filecolor=orange,      
    urlcolor=orange           
}
\usepackage{units}
\usepackage{color}

\newcommand{\scriptr}{\mathcalligra{r}\,}
\newcommand{\Mach}[0]{\mathcal{M}}

\def\orb{\rm{orb}}

\def\adot{\dot{a}}
\def\edot{\dot{e}}
\def\Mdot{\dot{M}}

\def\eg{{\em e.g.}}

\title[Eccentric Binaries in Retrograde Disks]{Eccentric Binaries in Retrograde Disks}

\author[Tiede $\&$ D'Orazio]{
Christopher Tiede$^{1}$\thanks{E-mail: christopher.tiede@nbi.ku.dk} \&
Daniel J. D'Orazio$^{1}$\thanks{E-mail: daniel.dorazio@nbi.ku.dk}
\\
$^{1}$Niels Bohr International Academy, Niels Bohr Institute, Blegdamsvej 17, 2100 Copenhagen, Denmark
}

\date{Accepted November 13, 2023. Received November 13, 2023; in original form July 10, 2023}

\pubyear{2024}

\begin{document}
\label{firstpage}
\pagerange{\pageref{firstpage}--\pageref{lastpage}}
\maketitle

\begin{abstract}
\noindent
Modern numerical hydrodynamics tools have recently enabled detailed examinations of binaries accreting from prograde circumbinary disks.
These have re-framed the current understanding of binary-disk interactions and disk driven orbital evolution. 
We present the first full-domain grid-based hydrodynamics simulations of equal-mass, eccentric binaries accreting from retrograde circumbinary disks.
We study binary eccentricities that span $e=0.0$ to $e = 0.8$ continuously, and explore the influence of retrograde accretion on the binary orbital response, disk morphology, and observational properties.
We find that, at all eccentricities, retrograde accretion shrinks the binary semi-major axis and pumps its eccentricity leading to the previously identified possibility of highly eccentric mergers.
Contrary to past studies and models, we observe gravitational forces to dominate the binary's orbital evolution as opposed to the physical accretion of mass and momentum.
Retrograde accretion variability also differs strongly from prograde solutions.
Preeminently, binaries with $e > 0.55$ reveal a unique two-period, double-peaked accretion signature that has not previously been identified.
We additionally find evidence for the emergence of retrograde Lindblad resonances at large eccentricities in accordance with predictions from linear theory.
Our results suggest that some astrophysical binaries for which retrograde accretion is possible will experience factors-of-a-few times faster orbital decay than in prograde disks and will have their eccentricities pumped beyond the limits found from prograde solutions. 
Such effects could lead to rapid inward migration for some young stellar binaries, the detection of highly-eccentric LISA mergers, and the tentatively observed turnover at the low-frequency end of the gravitational wave background.
\end{abstract}

\begin{keywords}
hydrodynamics -- software:simulations -- black hole mergers -- gravitational waves -- quasars:general
\end{keywords}
%


\section{Introduction} \label{S:Introduction}

Accretion onto a binary from a surrounding circumbinary disk (CBD) is important for the evolution and observation of many types of astrophysical binaries ranging from protoplanetary systems, to binary stars, to massive black hole binaries
\citep{KleyNelson:2012:rev, Orosz:2012Sci, Barnes:1996, Mayer:2013:MBHBGasRev}.
A plethora of work has been completed in recent years to determine the effect of prograde CBD's on the orbital evolution of the inner binary and on the associated observational characteristics \citep{MacFadyen:2008, Cuadra:2009, ShiKrolik:2012, DHM:2013:MNRAS, ShiKrolik:2015, MunozLai:2016, MirandaLai+2017, Yike+2017, Moody:2019, Tiede:2020, Duffell:2020, Zrake+2021, DOrazioDuffell:2021, Dittmann:2022, Franchini:2022, Siwek:2022}.
A primary result from the majority of these studies is that for the fiducial setup of an equal mass, circular binary embedded in an $\alpha$-disk \citep{SS73} with characteristic scale height $h/r = 0.1$ and turbulent viscosity parameter $\alpha = 0.1$, the CBD delivers angular momentum to the binary and drives binary outspiral \citep{Munoz:2019}.
The most recent studies have largely focused on subsequently filling out the parameter space beyond this reference model by varying the binary mass ratio \citep{Duffell:2020, Siwek:2022}, the binary eccentricity \citep{Zrake+2021, DOrazioDuffell:2021, Siwek:2023}, the disk scale-height \citep{Tiede:2020, HeathNixon:2020, Dittmann:2022, Penzlin:2022, Franchini:2022}, and the disk cooling \citep{Sudarshan:2022, WangBaiLai:2022, WangBaiLai:2023}.
Nearly all of these studies have found that both the direction and magnitude of binary inspiral depends non-trivially on each of these parameters; see \cite{LaiMunoz:Review:2022} for a recent review.

Many of these works have also detailed observational characteristics of these disks through measured accretion rates onto the binary as a proxy for  the emitted accretion flux \citep[\eg][]{Farris:2015:GW, Yike:2018}.
A primary result from prograde disks is that binaries accrete at a rate that is on average equal to that of a single central object. The underlying accretion modulations occur near the binary orbital frequency (and its harmonics) as well as on longer timescales at $\sim 5 \times$ the orbital period due to periodic over-feeding from a lopsided cavity formed in circular and near-circular binaries \citep{MacFadyen:2008, ShiKrolik:2012, DHM:2013:MNRAS, Farris:2014, Yike+2017, Duffell:2020, Dittmann:2022}.
Eccentric binaries in prograde disks, however, lose the longer $\sim 5$ orbit period and instead have their accretion variability dominated by the binary orbital frequency \citep{Hayasaki:2007, Dunhill+2015, MirandaLai+2017, Zrake+2021, Westernacher-Schneider:2022}.

Of primary relevance to this study, both \cite{DOrazioDuffell:2021} and \cite{Zrake+2021} found that equal-mass binaries that start with small eccentricity $e \lesssim 0.1$ have their eccentricity damped towards circular orbits---at which they are driven apart by the gas---but all other initial binary eccentricities $e \gtrsim 0.1$ are driven towards an equilibrium eccentricity $e_{\rm} \sim 0.4-0.45$.  
Further, \cite{Siwek:2023} demonstrated that this general behavior holds for all binary mass ratios $q > 0.1$, and the equilibrium eccentricity varies between $0.25 \lesssim e_{\rm eq} \lesssim 0.5$.
Such an equilibrium eccentricity could manifest in populations of stellar binaries that have undergone gas accretion phases as well as in residual eccentricity measurements of merging super-massive black hole binaries (SMBHB's) in gravitational waves with the space-based interferometer LISA.

Nearly all modern numerical studies have focused on prograde CBD's; but in some astrophysical systems it is not clear a priori what the angular momentum of the CBD ought to be relative to the binary.
In particular, SMBHB's that undergo active accretion phases in the late-stages of their evolution following a galaxy merger may receive gas injections isoptropically, and it is possible that retrograde gas feeding onto a SMBHB is comparably likely to prograde configurations \citep{KingPringle:2006, NixonKingPringle:2011, HobbsKing:2011}. 
It is also possible that a non-negligible fraction of young stellar binaries born in dense, chaotic star forming regions acquire retrograde CBD's (\eg, $\sim 10\%$ in \citealt{Elsender+Bate:2023}; \citealt{BateLodatoPringle:2010}).
This has motivated past numerical studies of retrograde-CBD's around circular binaries with viscous, smoothed particle hydrodynamics \citep{NixonKing_retro+2011, NixonLubow:RetroRes:2015} and inviscid 3D magnetohydrodynamics \citep{BankertShiKrolik:2015}. 
\cite{RoedigSG_retro:2014} have also simulated retrograde, self-gravitating disks around eccentric binaries for a few eccentricities up to $e=0.8$, and \cite{Amaro-Seoane_Retro+2016} examined the effect of different sink prescriptions for binaries with mass ratio $q=1 / 10$. 
These results have also motivated a series of approximate analytic models for retrograde binary accretion \citep{NixonKing_retro+2011, RoedigSG_retro:2014, Schnittman_Krolik_retro:2015, Amaro-Seoane_Retro+2016}. 
We review these works and their findings in more detail in \S\ref{S:LitReview}, but generally speaking, they all find that a retrograde-CBD causes accreting binaries at all eccentricities and mass ratios to shrink their separation (inspiral) and to pump their eccentricity (above some small threshold).

That said, the simulations performed for these studies were either focused on a few specific parameter setups or did not simulate the full multi-scale domain including both the entire CBD and the binary with its intra-orbit material\footnote{This has been demonstrated to be integral to converging on a solution in prograde scenarios \citealt{Farris:2014, Yike+2017, Munoz:2019}; with possible exceptions in confined regions of parameter space \eg, \citealt{Tiede:2022}.}.
Moreover, the developed analytic models of retrograde binary orbital evolution relied on these simulations and typically assumed the dominant contribution to be inelastic collisions with parcels of gas at binary apocenter.

In addition to being equally astrophysically viable as prograde solutions, retrograde circumbinary systems are analytically interesting because at low binary eccentricity they lack the Lindblad resonances that strongly affect---and possibly dominate \citep[see, however,][]{MaheshMcW+2023}---the prograde scenario \citep{GT79, GT80}.  Thus, retrograde CBD's offer an informative insight into the relevant dynamics of accretion onto eccentric binaries by comparison to their prograde counterparts.

In this paper we present the first full-domain, grid-based simulations of equal-mass binaries accreting from retrograde CBD's and expand upon recent works exploring the orbital evolution of eccentric binaries in prograde configurations -- this work can be directly compared with the recent prograde results of \cite{DOrazioDuffell:2021, Zrake+2021}.  \S\ref{S:LitReview} reviews the previous literature on retrograde binary accretion. \S\ref{S:Methods} lays out the numerical techniques used in this study. 
\S\ref{S:Results} includes our numerical results for binary evolution (\S\ref{s:adot-edot}), details of the disk morphology and how it changes with eccentricity (\S\ref{s:disk-morphology}), an analysis of the disk-mediated eccentricity evolution and its primary drivers (\S\ref{s:grav-v-acc}, \S\ref{s:ecc-driving}), the appearance of retrograde resonances (\S\ref{s:resonances}), and variability signatures from retrograde accretion (\S\ref{s:accretion_rate}).
We discuss how our results compare with previous works and explore observational implications in \S\ref{S:Discussion}, and summarize in \S\ref{S:Conclusions}.
The \hyperref[A:sinksoft]{Appendix} includes an additional study on how our results depend on the specifics of the sink prescription and gravitational softening.

\section{Retrograde Results in the Literature}
\label{S:LitReview}

The initial motivation for the study of retrograde CBD's was as an astrophysically plausible mechanism to shepherd SMBHB's from the canonical dynamical friction stalling radius \citep{Milosavljevic:2003:EvoMBHB} down to the sub-parsec separations where gravitational waves could merge the binary within a Hubble time. 
At the time, analytic and numerical works had suggested that prograde circumbinary disks would absorb angular momentum from the binary, becoming either decretion disks \citep{Webbink:1976, Pringle:1991} (that transfer mass outwards) or unstable \citep{Lodato:2009} halting binary migration.

The angular momentum transport in the prograde interaction was thought to be dominated by Lindblad resonances where the disk angular velocity $\Omega(r)$ was equal to the combination of binary orbital frequency $\omega_b$ and positive integer mode number $m = 1, 2, ...$ as $\Omega(r) = m \, \omega_b/(m \pm 1)$.   
However, this is only valid for $|\Omega| > |\omega_b|$ such that these resonances are not present when $\Omega$ and $\omega_b$ have opposite sign. \cite{NixonKing_retro+2011} explored the possibility that such retrograde CBD's could be a promising mode for shrinking an accreting SMBHB toward a GW dominated regime and coalescence.
They employ a toy model where the binary-disk interaction is assumed to only occur at apocenter and pericenter where the binary semi-major axis $a$ and eccentricity $e$ evolve assuming inelastic collisions between the smaller, secondary binary component of mass $M_2$ and Keplerian gas parcels of some mass $\Delta M$.
Their model finds that all binaries shrink their semi-major axis due to the loss of energy and angular momentum and note that eccentricity is pumped at binary apocenter and damped at pericenter.
As a result, binaries that start with small eccentricity tend to remain circular because of the relative balance between effects at pericenter and apocenter; but binaries that start with eccentricity greater than the characteristic scale-height of the disk $e \gtrsim h/r$ will have their evolution dominated by interactions at apocenter and have their eccentricity pumped until it is near unity and the binary coalesces.
They compare their toy model with a limited set of 3D viscous smoothed particle hydrodynamics (SPH) simulations and confirm that secondary-gas interactions at apocenter drive eccentricity and at pericenter damp eccentricity for a binary mass ratio of $q =1 / 10$.
They also include one simulation with $q=0.5$ and very small initial eccentricity and demonstrate that it stays near circular; but they do not include an example simulation of the case with initial eccentricity $e_0 > h/r$ where they would expect rapid eccentricity growth towards binary coalescence.
\cite{NixonKing_retro+2011} also point out that such an interaction ought to drive eccentricity in the interacting gas parcels and could potentially drive an eccentric CBD.

\cite{RoedigSG_retro:2014} simulated a more extensive set of eccentric binaries in retrograde CBD's using 3D SPH with $\beta$-cooling and self-gravity (instead of a typical viscosity). 
All of their simulations were for near-equal mass binaries with $q=1/3$, and similar to \cite{NixonKing_retro+2011}, they find that near-circular binaries in retrograde CBD's remain nearly circular while shrinking their semi-major axis (at a rate comparable to a prograde, $e=0$ control). 
For eccentricities above a critical value $e > 0.04$, they find that both $\{ \dot a, \, \dot e \}(e) \propto e$ such that eccentricity grows exponentially until the binary merges at eccentricities near unity.
Additionally, they amend the toy model of \cite{NixonKing_retro+2011} based on an impact-interaction at binary apocenter to include terms $\propto 1+q$ for when $q \ll 1$ is not satisfied and find an overall qualitative agreement; although they note that their model slightly underestimates both the binary semi-major axis shrinking and the eccentricity growth for $e \gtrsim 0.3$.

It is worth mentioning that one major discrepancy between the simulations of \cite{NixonKing_retro+2011} and \cite{RoedigSG_retro:2014} was that the former observed the presence of material around each binary component in the form of circum-single ``minidisks'', while the latter did not. \cite{RoedigSG_retro:2014} attributed this to their cooling prescription.

The first grid based simulations of retrograde circumbinary accretion were performed by \cite{BankertShiKrolik:2015} in 3D Newtonian MHD.  These simulations were for equal mass binaries $q=1$ fixed on a circular orbit where the grid had an inner boundary at $r = 0.8 a$ (with $a$ the binary separation) such that hydrodynamics in direct proximity of the binary components themselves were not resolved.
This study used time- and azimuthally-averaged radial torque density profiles from $r > 0.8 a$ to conclude that very little angular momentum is transferred gravitationally between the binary and the disk and that the dominant effect in the orbital evolution of the binary is from the direct accretion of mass and retrograde angular momentum; in general agreement with the modelling of \cite{NixonKing_retro+2011, RoedigSG_retro:2014}.
\cite{BankertShiKrolik:2015} also found the the retrograde CBD does not become eccentric and maintains its general axisymmetry unlike its prograde counterpart, but that the retrograde MHD disk around a circular binary does still exhibit spiral density perturbations despite the lack of standard Linblad and co-rotation resonances.

\cite{Schnittman_Krolik_retro:2015} used the conclusion from \cite{BankertShiKrolik:2015} that the binary evolution is dominated by the direct accretion of mass and momentum to develop an analytic model for binary separation and eccentricity evolution as a function of both binary eccentricity and mass-ratio.
This model was again based around the ``impact approximation'' that all energy and angular momentum exchange occurs at apocenter; however, their model is built around a value of the specific angular momentum exchange measured from \cite{BankertShiKrolik:2015} and includes the possibility of differential accretion between the two binary components.
Consistent with others, their model predicted semi-major axis shrinking for all eccentricities and mass ratios, but in contrast with previous modelling posited that eccentricity driving is maximal at small $e$ and decreases with growing eccentricity.

Using 2D viscous hydrodynamics, \cite{Amaro-Seoane_Retro+2016} studied the interaction between a binary with mass ratio $q=1/10$ and a retrograde CBD. 
They considered both circular setups and configurations with an initial eccentricity $e=0.6$.
They point out that because of the large relative velocities between the secondary and retrograde gas, it is important to use a sink radius---for the removal of gas at the location of the black hole (see Sec. \ref{S:Methods} for more precise definition and details)---that is smaller than the black hole's sphere of influence; and determine a sink radius of 2\% the secondaries Roche Lobe radius to be sufficient for a $q=1/10$ secondary.
Notably, they point out that this is because gas near to the secondary black hole can exert strong gravitational torques that alter the orbit, and too large a sink radius removes this gas from the domain resulting in erroneous estimates of the evolutionary effects.
Accordingly, they develop an analytic model for the evolution of binary semi-major axis and eccentricity for small mass ratios $M_2 \ll M_1$ based on both accretion and gas-dynamical friction from nearby, but unbound material. 
The outcome of this model depends on the steepness of the CBD density profile, but they find that most eccentric binaries will continue to have their eccentricity pumped $e \rightarrow 1$ by the retrograde CBD.
However, for near-circular binaries they find that relatively flat density profiles give eccentricity growth, but slightly steeper profiles with $\rho \propto r^{-n}$ and $n > 3/2$ near-circular binaries stay near-circular while shrinking their semi-major axis.
They find this model to be consistent with their circular and $e=0.6$ simulations, but do not perform an extensive comparison.

Each of the studies above has noted that because of the lack of binary-disk resonances, the retrograde CBD is not truncated away from the binary, but rather extends all the way down until it interacts directly with the binary orbit at apocenter.  
\cite{NixonLubow:RetroRes:2015} pointed out, however, that while this lack of Lindblad resonances is true for circular binaries, there exist components of the potential expansion for eccentric binaries that rotate retrograde to the binary allowing for resonant torques in eccentric, retrograde binaries. 
They find that these torques are generally weak, but for sufficiently high eccentricity, can become strong enough to drive spiral waves in the CBD and possibly carve a cavity (similar to prograde solutions).
They corroborate their analytic calculations with a suite of SPH simulations and find a critical eccentricity $e \sim 0.6$ where retrograde Lindblad resonances become comparable to viscous torques in the CBD for $\alpha$-viscosity value $\alpha=0.05$. They observe these resonances to drive spiral density waves into the CBD and to possibly carve a circumbinary-cavity (although these can be challenging to resolve in SPH simulations because of the extreme low-densities associated with circumbinary cavities).

\section{Numerical Methods}
\label{S:Methods}

For comparison and increased confidence in results, this study utilizes two separate grid-based Newtonian hydrodynamics codes: the moving-mesh code \texttt{Disco} in cylindrical coordinates \citep{DuffellMHDDISCO:2016}, and the GPU-accelerated code \texttt{Sailfish} in Cartesian coordinates \citep[see][]{Westernacher-Schneider:2022, Westernacher-Schneider:2023}. 
Both codes solve the the Navier-Stokes equations for viscous, locally isothermal hydrodynamics in 2-dimensions.
We refer the reader to an upcoming code-comparison paper \cite{Duffell:code-comparison} for implementation-specific details as well as a direct comparison of prograde CBD solutions.

%
\subsection{Problem setup} \label{s:setup}

We initialize the disk in two ways:
In \texttt{Sailfish}, the vertically-averaged surface density is initially uniform $\Sigma / \Sigma_0 = 1$ with a Keplerian rotation profile $v_\phi = \sqrt{GM / \scriptr}$. 
In \texttt{Disco}, a depleted central cavity of minimum density $\delta_0 = 10^{-5}$ is imposed on the otherwise uniform initial surface density profile
$\Sigma / \Sigma_0 = (1-\delta_0) e^{-(2.5/r)^{12}}  + \delta_0$
and the initial angular velocity of the gas is set by a Keplerian profile plus binary quadrupole and pressure corrections far from the binary, denoted $\Omega_0$ \citep[see, \eg,][]{MirandaLai+2017},
down to $r=a$ at which it saturates to $\Omega = \Omega_b \equiv \sqrt{GM / a^3}$ for $r < a$: 
    $\Omega(r) = \left[ \Omega_0(r)^{-4} + \Omega_b^{-4} \right]^{-1/4}$. 
As both sets of initial conditions capture steady state disk solutions far from the binary, and because we allow the disk to relax for $\gtrsim 500$ binary orbits before measurement (see \S~\ref{s:esweeping}), these small difference in set up do not have a significant impact on the results and diagnostics discussed below.

The disks are subject to the time-varying gravitational potential of a binary with separation $a$, orbital frequency $\Omega_b$, total mass $M$, and mass ratio $q$.
$\Sigma_0$ is an arbitrary density scaling (=1 in code units), and $\scriptr = \sqrt{r^2 + r_s^2}$ is the softened radial coordinate to prevent divergences at the origin (the softening radius is fiducially set to 5\% the binary separation, $r_s = 0.05\,a$).
The disk has constant kinematic viscosity $\nu = 10^{-3} \, a^2\Omega_b$ which induces radial inflow in the disk and implies a steady-state accretion rate at infinity $\dot M_0 = 3\pi \nu \Sigma_0$.

We consider the disk to be in the thin-disk limit with Mach number $\Mach = 10$ implying a disk aspect ratio $h/r \sim \Mach^{-1} = 0.1$. 
The locally isothermal condition is applied via the sound speed definition $c_s^2 = -\Phi_g / \Mach^2$ with $\Phi_g$ the binary potential
\begin{equation}
    \Phi_g = \Phi_1 + \Phi_2 \, , \qquad \Phi_j = -\frac{GM_j}{\scriptr_j},
    \label{eq:potential}
\end{equation}
and $\scriptr_j$ the smoothed distance to binary component $j \in \{1, 2\}$.
For this study we only consider equal mass binaries $M_1 = M_2 = M / 2$  (or equivalently, binary mass ratio $q=1$).
Moreover, we take the disk mass $M_d \sim \Sigma_0 a^2$ to be much smaller than the total binary mass $M_d \ll M$ such that the timescale for altering the binary orbit is long compared to the orbital time, and that we may ignore the disks self-gravity (\eg, the disk Toomre parameter $Q \sim \Mach^{-1}(M / M_d) \gg 1$).

%
\subsection{Source terms and boundary conditions} \label{s:sources}

The mass and momentum conservation equations solved in both codes can be written as
\begin{align}
    \partial_t \Sigma &+ \mathbf{\nabla} \cdot (\Sigma\mathbf{v}) = S_\Sigma \\
    \partial_t (\Sigma \mathbf{v}) &+ \mathbf{\nabla} \cdot (\Sigma \mathbf{v} \mathbf{v} + P\mathbf{I} - \mathbf{\tau}) = \mathbf{S_J} + \mathbf{F_g}
\end{align}
where $v_i$ is the mid-plane fluid velocity, $P = c_s^2 \Sigma $ is the vertically averaged gas pressure, $\tau$ is the viscous stress tensor, $F_g$ is the gravitational force density from the potential in Eq. \ref{eq:potential}, and $S_{\{\Sigma, J\}}$ are mass- and momentum-sinks respectively.
The viscous stress tensor is calculated as 
\begin{equation}
    \tau_{ij} = \nu \Sigma (\nabla_i v_j + \nabla_j v_i - \delta_{ij} \nabla_k v_k) 
\end{equation}
with the covariant derivatives taken in the relevant coordinate system.
The sink terms are included to mimic the accretion of material onto each binary component because we do not resolve our solutions down to the physical accretion boundaries.
The mass and momentum sinks are given respectively as
\begin{align}
    S_\Sigma     &= -\gamma \Omega_b \Sigma \sum_i w_i \\
    \mathbf{S_J} &= -\gamma \Omega_b \Sigma \sum_i \mathbf{v}_i^\star \,w_i
\end{align}
where $\gamma$ is a dimensionless sink-rate which---unless specified otherwise---is set to 1, and $\Omega_b = \sqrt{GM / a^3}$ is the Keplerian orbital frequency of the binary. 
$w_i$ is a window function defining the strength of the sink for a gas parcel distance $r_i$ away from black hole $i$ in terms of some characteristic sink radius that is fiducially set equal to the softening radius $r_s$:
\begin{align}
    w_i = \exp{\left[ -( r_i / r_s )^4 \right]} \ .
\end{align}

Different sink behaviors are achieved through calculation of the adjusted gas-velocity $\mathbf{v_i^\star}$ associated with each  fluid element removed.  
For nearly all runs presented in this study we have adopted ``torque-free'' \citep{Dempsey-TFsinks:2020}---or \emph{spinless}---sinks such that no spin angular momentum is accreted by each binary component---only orbital angular momentum.
To accomplish this, the adjusted velocity is calculated as
\begin{equation}
    \mathbf{v_i^\star} = (\mathbf{v} - \mathbf{v_i}) \cdot \mathbf{\hat r_i} + \mathbf{v_i},
\end{equation}
where $\mathbf{v}$ is the gas velocity in the inertial frame, $\mathbf{v_i}$ is the binary component velocity, and $\mathbf{\hat r_i}$ is the radial unit-vector in a coordinate system centered on binary component $i$.
In the appendix we briefly discuss a separate ``acceleration free'' sink prescription (also sometimes referred to as a ``standard sink'' \citealt{Dittmann:2021}) in which $\mathbf{v_i^\star} = \mathbf{v}$ and the momentum sink reduces to $\mathbf{S_j} = S_\Sigma \mathbf{v}$.

In both codes the outer boundary is set so to enforce the steady-state accretion rate $\dot M_0$ so that the solution mimics that of an infinite accretion disk.
In \texttt{Disco} this is accomplished with outer ghost-cells fixed to the setup initial conditions.
In \texttt{Sailfish} this is attained via an additional ``buffer'' source term that drives the solution back towards the initial condition. 
This ``buffer'' prevents artifacts from the square domain propagating into the solution.
The details of the buffer prescription can be found in \S3 of \cite{Westernacher-Schneider:2022}.

%
\subsection{Adiabatic eccentricity variation} \label{s:esweeping}

The primary calculations presented in this study fix the binary mass ratio $q=1$ and slowly vary the binary eccentricity from $e_0=0$ up to $e_f=0.8$.  
In the process we measure the rate of change of both the binary semi-major axis $\dot a (e)$ and eccentricity $\dot e (e)$ as in \citet{DOrazioDuffell:2021} (see also \citealt{Duffell:2020} for the same procedure performed on $q$ at fixed-$e$, and \citealt{WangBaiLai:2022} on the disk cooling timescale).
The runs that vary the binary eccentricity are initialized from the output of a simulation with eccentricity fixed at $e=0$ for 500 binary orbits.
The eccentricity growth is performed linearly over $n$ binary orbits as
\begin{equation}
    e(t) = e_0 + \frac{e_f}{2\pi\,n}\,t \, .
    \label{Eq:esweep}
\end{equation}
We take $n = 5 \times 10^3 \,\text{and}\, 10^4$ and have verified that further increasing $n$ does not meaningfully change our results. The time rate of change of the binary orbital elements $\dot a$, $\dot e$ are calculated in the same way as is detailed in \cite{DOrazioDuffell:2021}, except with additional full accounting for all momentum accreted by the sinks (the mass deposition effect is included as previously).

The fiducial grid resolution in \texttt{Sailfish} is taken to be $\delta = 0.0067 a$. 
In \texttt{Disco}, the grid is hybrid-logarithmic in radius such that the resolution is $\delta = 0.0127 a$ at $r=0.5\,a$, slightly higher inside $r < 0.5\,a$, and decreases as one moves toward the outer boundary at $r=50a$.
In \texttt{Sailfish}, the resolution is constant everywhere and the outer boundary is taken to be $r=10a$ because initial simulations (and the previous numerical work discussed in \S\ref{S:LitReview}) showed that the retrograde CBD becomes near completely axisymmetric at $r > \text{few} \times a$, with possible exception at large eccentricity to be discussed later.

\section{Numerical Results}
\label{S:Results}
%

%
\subsection{Orbital evolution: $\dot{a}(e)$ and $\dot{e}(e)$} 
\label{s:adot-edot}
%

\begin{figure*}
  \includegraphics{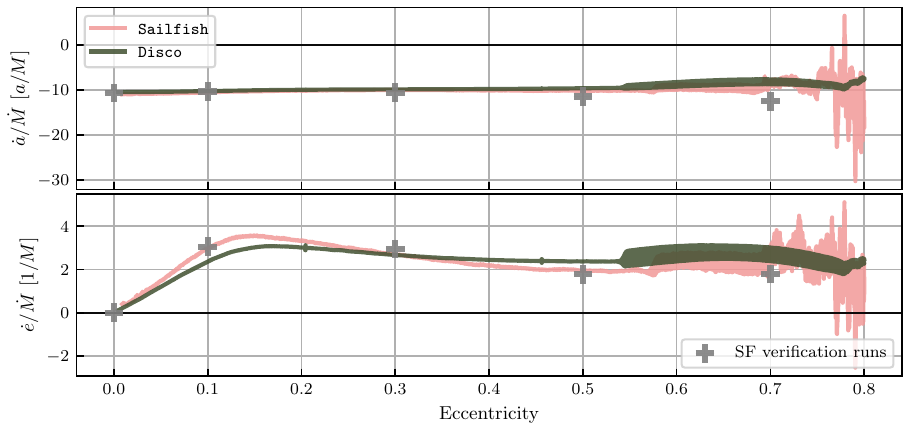}
  \vspace{-5pt}
  \caption{Time rates of change of binary semi-major axis (\emph{top}) and eccentricity (\emph{bottom}) from a retrograde circumbinary disk as a function of binary eccentricity. 
  Results from \texttt{Sailfish} are shown in pink and from \texttt{Disco} in green.
  The grey crosses show single, fixed-eccentricity runs computed with \texttt{Sailfish} for comparison to the results from the eccentricity-varying runs.}
 \label{fig:adot_edot}
\end{figure*}

Figure \ref{fig:adot_edot} shows calculations of $\dot a(e)$ (\emph{top panel}) and $\dot e(e)$ (\emph{bottom panel}) from the simulations that vary eccentricity adiabatically from $e=0$ to $e=0.8$. 
Results from \texttt{Disco} are displayed in green and results from \texttt{Sailfish} are given in pink.
Both codes find that the retrograde CBD shrinks the binary semi-major axis $\dot a < 0$ and pumps the binary eccentricity $\dot e > 0$ for all eccentricities.
The shrinking of the binary orbit and pumping of the binary eccentricity is in general agreement with previous studies of binary evolution in retrograde CBD's, but unlike \cite{NixonKing_retro+2011} and \cite{RoedigSG_retro:2014} these simulations show that near-circular binaries are not in fact driven back towards circularity, but rather have their eccentricity driven up by the retrograde CBD.
The eccentricity driving rate grows linearly towards a maximum of $de / d\log{M} \approx 3$ at $e \approx 0.15$ and settle to a near constant growth rate $de / d\log{M} \approx 2.25$ by $e \gtrsim 0.4$.
The shrinking of the binary $\dot a$ shows almost no dependence on binary eccentricity and always retains a value $d\log{a} / d\log{M} \approx -10$.
\footnote{Curiously, for a circular binary $d\log{a} / d\log{M} \approx -10$ corresponds to an accretion eigenvalue (see \citealt{LaiMunoz:Review:2022}) $\ell_0 = \dot L / \dot M = -1$.
This value, however, is likely a coincidence because it remains constant with binary eccentricity when changes to the semi-major axis are no longer defined only by the flow of angular momentum in the disk.
}

To sanity check these results we have also run a series of high-resolution $\delta = 0.005\,a$ fixed-eccentricity runs in \texttt{Sailfish} shown as grey crosses. 
These runs were done for 2000 binary orbits and we report the rate of change to the orbital elements over the final 500 orbits.
These show near exact agreement with the \texttt{Sailfish} variable eccentricity run and very close agreement with the \texttt{Disco} run.
At high eccentricities $e > 0.7$ we observe a notable growth in the fluctuations of $\dot a$ and $\dot e$ in the \texttt{Sailfish} runs alone.
We attribute this growth in variation in one code and not the other to the fact that the Cartesian grid of \texttt{Sailfish} evolves the fluid linear momenta, and thus, angular momentum is not perfectly conserved.
\texttt{Disco} by contrast explicitly evolves and conserves the fluid angular momentum by construction.
We have confirmed this effect by lowering the resolution of the \texttt{Sailfish} run and verifying that the variations become more extreme, that they set in at lower eccentricity, and that the solutions begin to diverge at large eccentricities.

Apart from these variations, both simulations show a notable increase in signal variability at $e \sim 0.55$. 
We attribute this growth in variability to the emergence of a two-orbit periodic switching in the circumbinary disk structure (discussed further in \S \ref{s:disk-morphology}).
This growth in variability also appears to be coincident with the emergence of two-armed spiral density waves that extend from the CBD cavity edge suggesting the possible realization of retrograde circumbinary resonances in accordance with the predictions of \cite{NixonLubow:RetroRes:2015}.

%
\subsection{Disk morphology} \label{s:disk-morphology}
%

\begin{figure}
  \centering
  \includegraphics{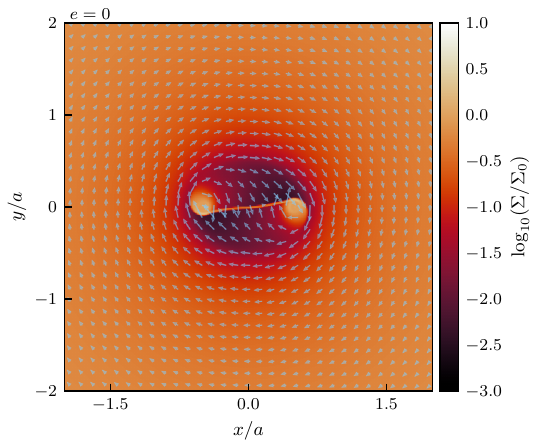}
  \vspace{-15pt}
  \caption{Snapshot of the steady-state flow pattern for a circular binary in a retrograde CBD. The arrows show the direction of the fluid velocity. The binary is orbiting counter-clockwise. The minidisks are retrograde in accordance with the bulk flow. The minidisks have persistent wakes that also feed low-angular momentum material into the standing-bridge between the binary components.}
  \label{fig:e00_snapshot}
\end{figure}

\begin{figure*}
  \centering
  \includegraphics{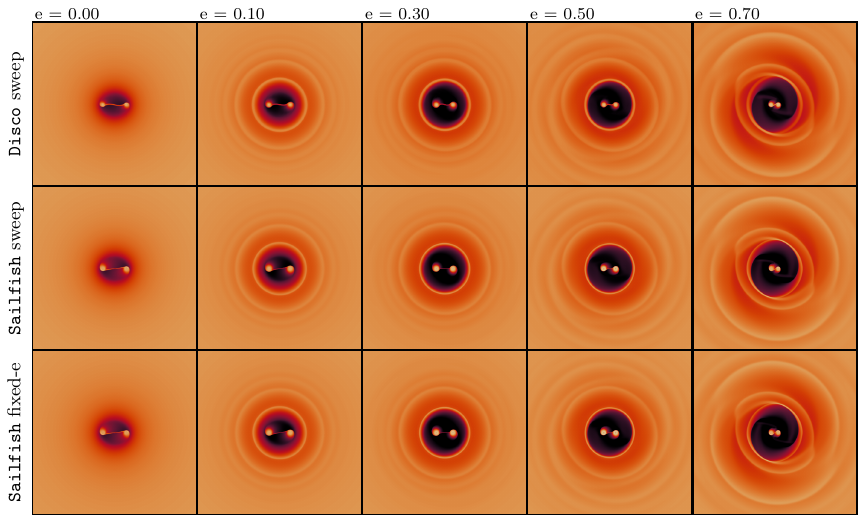}
  \caption{Density maps at $e=\{0.0, 0.1, 0.3, 0.5, 0.7 \}$ from the eccentricity varying runs with both \texttt{Disco} (\emph{top row}) and \texttt{Sailfish} (\emph{middle row}) as well as the fixed-eccentricity runs (\emph{bottom row}), all taken at pericenter.}
 \label{fig:sigma_maps}
\end{figure*}

A single snapshot for a circular binary ($e=0$) after 2000 orbits is shown in Figure \ref{fig:e00_snapshot} that highlights many of the general features of retrograde solutions.
The colormap shows the log-density of the accretion flow, and the overlayed arrows show the direction of the gas velocity.
Similar to previous studies, the circumbinary disk extends all the way to the binary orbital radius, and remains almost completely axisymmetric everywhere except for the inner-most $r < a$ (in contrast with the large, lopsided cavities observed in circular, prograde scenarios).
The binary carves a low-density cavity inside of its orbital radius, but does not entirely expel it of material, and rather is always orbitting in some ambient medium.
The binary components capture retrograde material into circum-single ``minidisks'' that retain opposite angular momentum to the orbital angular momentum of the binary.
Rather than becoming perfectly circular, the minidisks have bow-shock-like structures on their leading edge from the ram-pressure of incident counter-rotating material, as well as trailing ``tadpole wakes'' of low angular momentum material that falls almost radially toward the binary barycenter in both directions. This creates a persistent standing ``retrograde-bridge'' of material between the binary components.
This retrograde-bridge is quasi-steady---it does wobble slightly---in circular retrograde solutions, but becomes more transitory for eccentric binaries as will be demonstrated later.
We also note that all future surface density plots will use the same colorbar indicated in Figure \ref{fig:e00_snapshot} and the same $[-2a, 2a]$ Cartesian stretch (unless otherwise indicated).

\begin{figure*}
  \centering
  \includegraphics{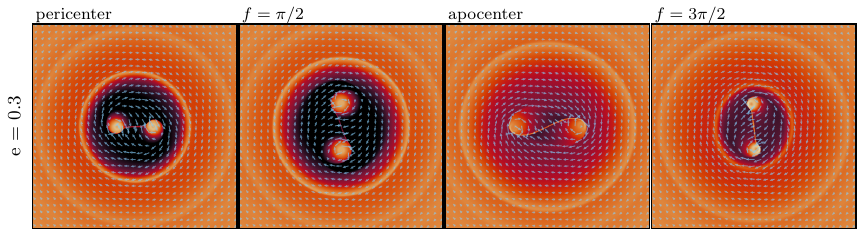}
  \caption{Phase dependent disk morphology for an $e=0.3$ orbit at four values of the true anomaly $f = \{0, \pi/2, \pi, 3\pi / 2\}$. Overlayed arrows show the direction of the fluid velocity.}
  \label{fig:e03_tseries}
\end{figure*}

\begin{figure*}
  \centering
  \includegraphics{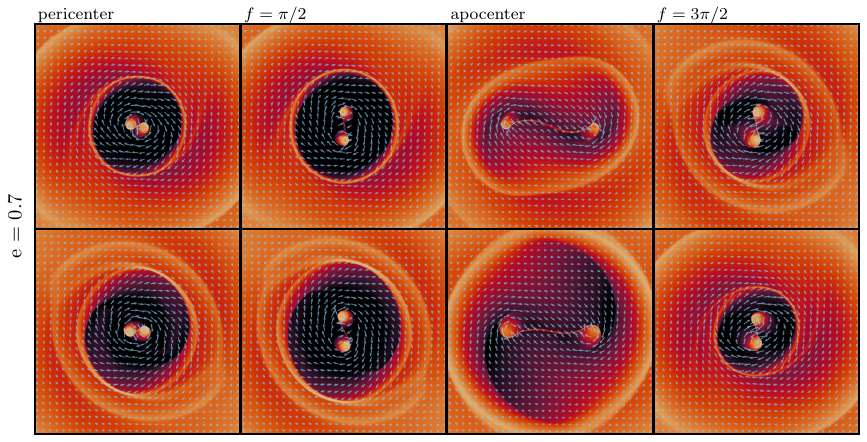}
  \caption{Phase dependent disk morphology for two contiguous $e=0.7$ orbits at four values of the true anomaly showing the two-orbit periodicity in the flow behavior.
  The first row illustrates the first orbit in this two-orbit behavior, and the second row shows the second.
  Overlayed arrows show the direction of the fluid velocity.}
  \label{fig:e07_tseries}
\end{figure*}

Figure \ref{fig:sigma_maps} displays snapshots of the disk $\log$-surface-density for eccentricities $e = \{0.0, 0.1, 0.3, 0.5, 0.7\}$ with Cartesian extent $[-3a, 3a]$. The \emph{top two rows} sample from the eccentricity varying runs while the \emph{bottom row} uses results of the fixed-eccentricity runs.
All snapshots are taken at binary pericenter.
There is generally very good agreement in disk morphology between the eccentricity varying runs (\texttt{Disco} sweep and \texttt{Sailfish} sweep; \emph{rows 1 and 2}) and with the fixed-eccentricity runs (\emph{row 3}).
The only discernible differences are in slight variations to the minidisk density and the exact shape of the retrograde-bridge between the binary components. We investigate the sensitivity of these wakes and bridges to numerical choices in \S~\ref{A:sinksoft}.

The time-dependence of the disk evolution has three separate regimes: (\emph{i}) steady-state in the co-rotating frame near circularity, (\emph{ii}) the driving of axisymmetric density waves for eccentric binaries $0.025 \lesssim e \lesssim 0.55$, and (\emph{iii}) the forcing of non-axisymmetric spiral density waves at large eccentricities $e \gtrsim 0.55$.
Circular binaries (\emph{i}) with $e \lesssim 0.025$ resemble Figure \ref{fig:e00_snapshot} and show no phase dependence in their evolution.
The systems in regime (\emph{ii}) acquire a ``breathing-mode'' that drives axisymmetric density waves into the CBD.
This process is shown in Figure \ref{fig:e03_tseries}.
At pericenter, the binary has carved a fully-depleted, axisymmetric cavity.
As it approaches apocenter (true anomaly $0 < f < \pi$; here-on we refer to this as ``binary waxing''), the minidisks circularize in the relative-vacuum of the cavity, but CBD material encroaches on the cavity as the orbit expands and tidally redirects material through the domain center.
This redirected material from each component collides forming the retrograde-bridge, and by binary apocenter the cavity has been replenished with low-density gas.
The ram-pressure of this ambient material disrupts the minidisks compressing the leading edge and stripping off a spiral wake.
As the binary accelerates towards pericenter ($\pi < f < 2\pi$; ``binary waning''), it expels the material inside of its orbit driving an axisymmetric density ring into the CBD.
This ring can be seen forming in the final panel of Figure \ref{fig:e03_tseries} ($f = 3\pi / 2$), as circularized in the first panel (at pericenter), and propagating through the CBD as a sound wave in the second and third panels.
We illustrate $e=0.3$ as an example, but all eccentricities $0.025 \lesssim e \lesssim 0.55$ were observed to have the same phase-dependent behavior.
We note that the driving of an axisymmetric density wave sets in at very small eccentricity $e \approx 0.025$, but the carving of a depleted cavity around pericenter, however, occurs more gradually.
A fully-depleted cavity is effectively formed once per orbit starting at $e \approx 0.1$.

Binaries in regime (\emph{iii}) become qualitatively different as they acquire a two-orbit periodicity in the phase-dependence of their flow.
These two orbits are illustrated in Figure \ref{fig:e07_tseries}.
Similar to regime (\emph{ii}), at the first pericenter (selected arbitrarily, but shown at the top-left panel of Figure \ref{fig:e07_tseries}) the binary has carved a fully-depleted, mostly axisymmetric cavity with the exception of two weak spiral arms extending from the cavity wall.
In the first orbit (\emph{top row}), during binary waxing, the minidisks similarly circularize and material is tidally re-oriented into the depleted cavity.
At apocenter, the cavity is replenished with gas, the retrograde-bridge has formed, and the minidisks are strongly perturbed from the collisions with encroaching CBD material.
In the approach to pericenter, the accelerating binary again expels gas from within its orbit, but instead of driving an axisymmetric density wave (as in the lower $e$ case), it propels two spiral density waves that propagate into the disk.
Because of this, at second-pericenter (bottom left panel), the binary has carved a depleted cavity, but the cavity wall is no longer circular as it is dominated by the $m=2$ spirals.
In the second orbit (\emph{bottom row}), the same processes occur, but the non-axisymmetric cavity at pericenter is less efficient at refilling the cavity during binary waxing.
The minidisks are less-perturbed at second-apocenter (in the bottom row) as a result of these lower cavity densities; and while the binary still creates a two-armed spiral structure upon expelling its intra-orbit material during binary waning, the lower densities weaken the response, and the resulting pericenter cavity is left essentially circular.
This two orbit periodicity also appears as power in the accretion rate time series and periodogram (see Figures \ref{fig:2Dpdgm} and \ref{fig:Mdots_of_e}).

We emphasize that at all eccentricities, these 2D Newtonian, isothermal hydrodynamics simulations show the formation of binary minidisks and a retrograde-bridge.
The presence of these minidisks---and their visible asymmetries throughout the binary orbit---preempt the importance of including the binary and the central most $r < a$ regions of the accretion flow in order to accurately ascertain the gravitational forces on the binary and its resultant orbital evolution.

%
\subsection{Gravity vs. accretion} \label{s:grav-v-acc}

\begin{figure}
  \includegraphics{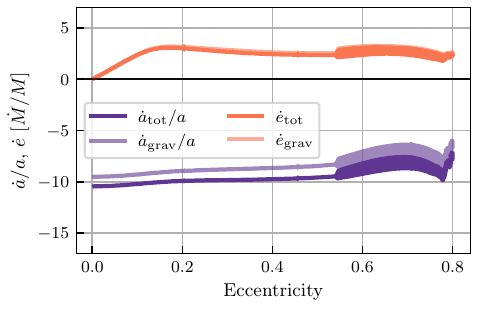}
  \includegraphics{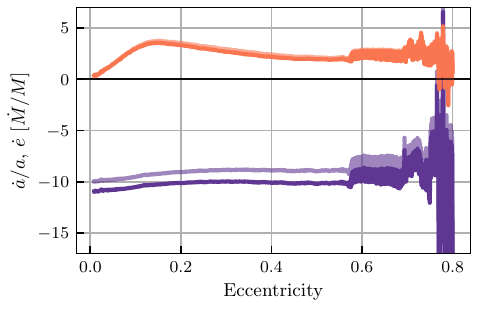}
  \vspace{-5pt}
  \caption{Orbital evolution from \texttt{Disco} (\emph{top}) and \texttt{Sailfish} (\emph{bottom}) separated into the total effect and the component from gravitational forces alone. The semi-major axis and eccentricity evolution are both dominated by the gravitational pull of the gas.}
 \label{fig:decomposition}
\end{figure}

A major component of most previous studies of orbital evolution from retrograde CBD's was that the evolution is dominated by the direct accretion of both mass and angular momentum due to collisions between retrograde fluid elements and the binary at apocenter; 
with the exception of \cite{Amaro-Seoane_Retro+2016} who pointed out that gas near the binary orbit can exert very strong gravitational forces before physically accreting.
In order to quantify the relative effects of gravitational and accretion forces, we separated both $\dot a$ and $\dot e$ into their components due solely to gravitational forces and those due only to the accretion of mass and momentum. 
Figure \ref{fig:decomposition} shows this decomposition by plotting the total $\dot a$ (\emph{purple}) and $\dot e$ (\emph{orange}), and the components due to gravitational forces alone $\dot a_{\rm grav}$, $\dot e_{\rm grav}$ (illustrated as \emph{light-purple} and \emph{light-orange}).
The top panel shows this decomposition for the \texttt{Disco} run and the bottom panel shows it for the \texttt{Sailfish} simulation.
The primary conclusion from these breakdowns is that---contrary to previous studies and models---gravitational forces are the dominant component of the binary's orbital evolution for both $\dot a$ and $\dot e$ as their components due to gravity alone near perfectly describe the full orbital evolution curves.

The effect from the physical accretion of mass and momentum is almost entirely negligible for the eccentricity evolution in both codes and only represents a $\sim 10\%$ contribution to the change in semi-major axis.
We posit that this---just as was pointed out in \cite{Amaro-Seoane_Retro+2016} for $q=1/10$ binaries---is because material captured from the CBD at binary apocenter is not directly added to the binary via accretion, but rather, is transferred onto orbits around the individual binary components.
Moreover, these circum-single, ``minidisk'' structures are not symmetric around each binary component, but instead have small wakes that trail each component exerting gravitational torques on and removing energy from the binary orbit.
Even larger wakes and non-symmetric features are formed when the minidisks are partially disrupted from the impact with the CBD at each apocenter passage as discussed in \S\ref{s:disk-morphology}.

\begin{figure}
 \centering
  \includegraphics{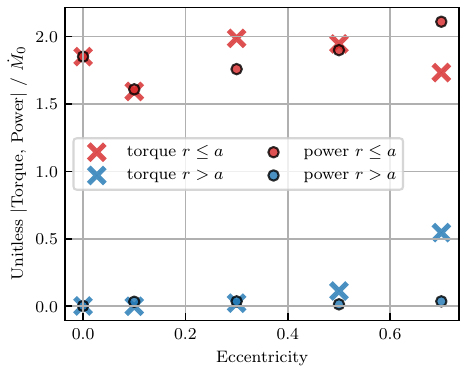}
  \vspace{-5pt}
  \caption{Magnitude of gravitational torque and power for fixed-eccentricity runs with \texttt{Sailfish}. We see that gravitational forces from the inner-most region of the flow dominate the binary evolution at all eccentricities.  At high-eccentricities $e \gtrsim 0.5$, the binary drives spiral wakes into the outer-CBD, leading to growth in the torque component from $r > a$; although this component remains subdominant by a factor of a few.}
 \label{fig:torque_power}
\end{figure}

To compare the extent to which gravitational forces from this intra-orbit material (material within $r \leq a$) dominate over those from the outer CBD ($r > a$), Figure \ref{fig:torque_power} shows the average magnitudes of the unit-less torque $|(\mathbf{r_b} \times \mathbf{f_g}) / \sqrt{GMa(1 - e^2)}|$ and power $|(\mathbf{f_g} \cdot \mathbf{v_b}) / (GM / 2a)|$, with $r_b$ the binary separation,
exerted on the binary from each of the fixed-eccentricity comparison runs.
The average torque is shown by crosses and the average power by circles.
The components from the intra-orbit material with $r \leq a$ are given in red and those from the outer CBD in blue.
At all eccentricities both the gravitational torque and power exerted on the binary are dominated by the intra-orbit material.
Further, the components of the average torque and power from the outer-CBD $r > a$ are almost entirely negligible at eccentricities below the threshold for retrograde resonances to drive spiral density waves into the CBD, $e \lesssim 0.55$.
At $e = 0.7$ we once again see the effect of persistent spiral density waves in the CBD as the average torque from the outer-disk is no longer negligible---and instead accounts for approximately 25\% of the total torque on the binary.

%
\subsection{Eccentricity driving} \label{s:ecc-driving}
%

\begin{figure}
 \centering
  \includegraphics{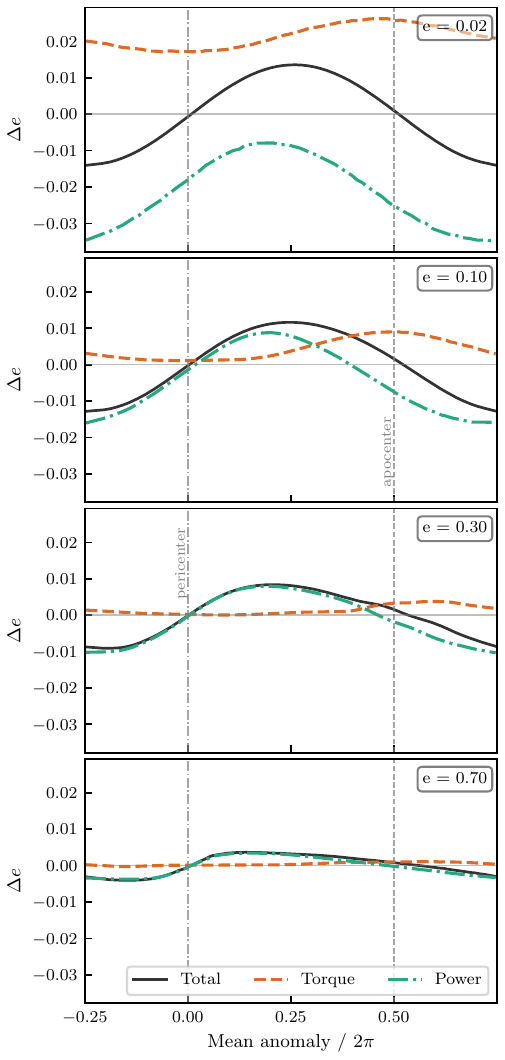}
  \vspace{-15pt}
  \caption{Averaged change in eccentricity $\Delta e$ per binary phase (mean anomaly) at four eccentricities $e = \{0.02, 0.1, 0.3, 0.7\}$. The top panel---representative of small eccentricities $e \sim \mathcal{O}(10^{-2})$ in a \emph{wake-driven} mode---shows different phase-dependent eccentricity driving behavior than all other eccentricities $e \gtrsim 0.1$ in a 
  \emph{cavity-mediated} mode.}
 \label{fig:edot_v_phase}
\end{figure}

\begin{figure}
 \centering
  \includegraphics{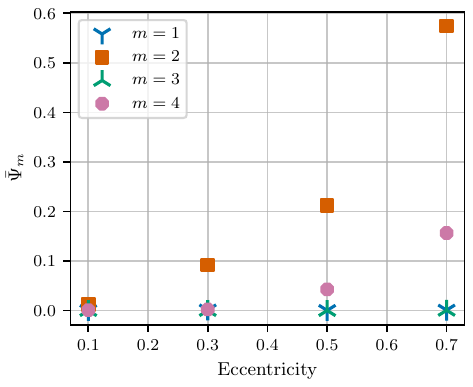}
  \vspace{-10pt}
  \caption{Time averaged strength of azimuthal density modes with mode numbers $m = \{1, 2, 3, 4\}$. We see the emergence of only even-$m$ azimuthal modes in accordance with linear theory of binary-disk resonances for equal mass binaries.
  The strong growth of the $m=2$ mode at $e = 0.7$ reflects the observed $m=2$ spiral density waves in the associated surface density plots and is evidence for the emergence of retrograde resonances.
  The non-zero even-$m$ mode strengths at $e < 0.55$ are likely due to the temporary non-axisymmetry of the CBD around binary apocenter once every orbit.}
 \label{fig:moments}
\end{figure}

A primary component of past modelling for binaries accreting from retrograde disks was the assumption that eccentricity pumping is focused around binary apocenter (and damping---if included---occurs around pericenter).
We examine this hypothesis by plotting the average change in binary eccentricity 
$\Delta e  (\theta) = \langle \dot e \delta t \rangle_\theta$ at each binary phase (orbital mean anomaly, $\theta$) in Figure \ref{fig:edot_v_phase} with $\dot e$ the instantaneous rate of change, $\delta t$ the timestep, and $\langle \rangle_\theta$ denoting the average at mean anomaly $\theta$.
$\Delta e$ includes the fact that the binary spends different amounts of time at each orbital phase, and integrating the curve over all phases would yield the average per-orbit change in eccentricity.
This tells us the relative contribution---on average---of each binary phase to the net change in binary eccentricity per orbit.
We show the total effect from both gravitational forces and the accretion of mass and momentum as solid black curves as well as the decomposed contributions from gravitational torque (\emph{orange dashed} curves) and gravitational power (\emph{green dash-dotted} curves) given respectively as 
\begin{align}
    \dot e_{\rm power} = -\frac{\dot \epsilon}{\epsilon} \left( \frac{1 - e^2}{2e} \right) &= \frac{\mathbf{f_g} \cdot \mathbf{v_b}}{GM / a} \left( \frac{1 - e^2}{e} \right) \\
    \dot e_{\rm torque} =  -2 \frac{\dot \ell}{\ell} \left( \frac{1 - e^2}{2e} \right) &= - \frac{\mathbf{r_b} \times \mathbf{f_g}}{\sqrt{GMa(1 - e^2)}} \left( \frac{1 - e^2}{e} \right),
\end{align}
such that $\dot e_{\rm grav} = \dot e_{\rm power} + \dot e_{\rm torque}$
(see \citealt{DOrazioDuffell:2021} Equations 6-8 for more detailed discussion of these terms).
These phased eccentricity driving curves are shown 
for four values of eccentricity $e = \{0.02, 0.1, 0.3, 0.7\}$ to encapsulate each of the three morphologic disk regimes: $e=0.02$ in regime (\emph{i}), $e=0.1,\, 0.3$ in regime (\emph{ii}), and $e=0.7$ in regime (\emph{iii}) (see \S \ref{s:disk-morphology}).
Of primary note, eccentricity pumping is found to occur during binary waxing (pericenter $\rightarrow$ apocenter) for all eccentricities, peaking near a mean anomaly of $\pi / 2$ (or equivalently at $P_b / 4$) at small eccentricities and shifting nearer to pericenter with growing $e$. 
Eccentricity is correspondingly damped during binary waning (apocenter $\rightarrow$ pericenter), and the effect at pericenter and apocenter is minimal at all eccentricities considered.
The net integrated effect over one full orbit yields the $\dot e_{\rm grav}$ curves shown in Figure \ref{fig:decomposition}.

One major distinguishing characteristic between regime (\emph{i}) and regimes (\emph{ii}, \emph{iii}) is that in regime (\emph{i}) the binary is always orbiting through ambient intra-orbit material and as a result forms persistent tadpole wakes that trail each binary component.
In regimes (\emph{ii} \& \emph{iii}), the binary expels all intra-orbit material once per orbit during its waning phase such that it evolves in a fully-depleted cavity for at least half its orbit until material is once again redirected within the orbital radius around apocenter (see Figures \ref{fig:e03_tseries} \& \ref{fig:e07_tseries}).
We identify two separate eccentricity driving modes associated with these disk morphologic regimes:
(1) a \emph{wake-driven} mode when the binary evolves through persistent intra-orbit material and never carves a fully depleted cavity (regime \emph{i}; see Figure \ref{fig:e00_snapshot}).
The associated eccentricity driving is shown in the top-panel of Figure \ref{fig:edot_v_phase} for $e=0.02$ where torques from the tadpole wakes pump eccentricity at all binary phases.
The associated power from the wakes always acts with opposite effect, damping the binary's eccentricity; but the torque-mediated pumping wins out on aggregate.

The second eccentricity driving mode, (2) a \emph{cavity-mediated} mode, occurs once the binary has begun carving a fully depleted cavity and orbits in relative vacuum for a significant portion of its orbit (regimes \emph{ii} \& \emph{iii}). 
This mode is observed in the $e = 0.1,\, 0.3,\, 0.7$ panels of Figure \ref{fig:edot_v_phase} where the total effect on $\Delta e$ is dominated by, and near perfectly tracks, the power contribution.
Once the binary begins carving a depleted cavity, gravitational power continues to damp eccentricity during binary waning as the binary expels its intra-orbit material.
However, the power effect switches to eccentricity pumping during waxing when the binary and its minidisks evolve through relative vacuum.
This transition occurs gradually, but has mostly set in by $e \approx 0.1$.
The effect of torque on $\Delta e$ during this mode is negligible at all phases, except near apocenter when the binary has refilled its cavity and temporarily evolves through ambient material with the associated tadpole wakes.
This effect, though, becomes less significant with growing eccentricity.

We attribute the initially small eccentricity driving at small eccentricities $e \sim \mathcal{O}(10^{-2})$ in Figure \ref{fig:adot_edot} to the small net difference between the eccentricity changing effects of gravitational torque and power from the wakes before the binary begins to carve a depleted cavity once per orbit.
The slow linear growth in $\dot e$ with $e$ reflects the gradual transition to a cavity-carving binary; and once the binary is effectively depleting its orbit of material around pericenter---despite the amplitude of all eccentricity modifying effects decreasing with $e$---the net balance remains relatively constant with $e$. 

%
\subsection{Retrograde resonances} \label{s:resonances}

As discussed in \S\ref{S:LitReview}, \cite{NixonLubow:RetroRes:2015} pointed out that eccentric binaries have components of the potential expansion that rotate retrograde to the binary admitting retrograde Lindblad resonances; and their simulations find that these resonances become strong enough to drive persistent spiral density waves into the disk at eccentricities $e \geq 0.6$.
The simulations presented in this study also suggest evidence for the emergence of retrograde resonances at eccentricities $e \gtrsim 0.55$.
Figure \ref{fig:sigma_maps} shows the first signs of non-axisymmetric density waves at $e=0.5$, and at $e=0.7$ reveals the presence of two-armed, $m=2$ spiral density waves originating at the CBD inner edge and propagating into the CBD.

In traditional linear analyses of resonant torques on a CBD, the forces associated with a harmonic mode $m$ of the gravitational potential will drive spiral density waves into the CBD with the same azimuthal mode number. 
Moreover, for equal-mass binaries, odd components of the potential expansion disappear, so only even-$m$ resonant torques should act on the CBD\footnote{and the presence of odd azimuthal modes---the $m=1$ mode in particular---has been suggested as evidence for non-resonant effects in the cavity-carving process for equal mass binaries \citep{MaheshMcW+2023}.}.
To check, albeit indirectly, for the presence of such resonances, we calculate timeseries of the azimuthal density mode $m = \{1,2,3,4\}$ as 
\begin{equation}
    \Psi_m(t) = \int_a^{r_{\rm out}} dr\, \int_0^{2\pi} r d\phi\, \Sigma(t)\,e^{im\phi}   .
\end{equation}
Figure \ref{fig:moments} shows the time averaged strength of these azimuthal density modes $\bar \Psi_m$ from the fixed-eccentricity runs.
We observe the presence of an $m=2$ azimuthal mode consistent with the observation of $m=2$ spiral density waves at $e > 0.55$.
However, this mode appears to emerge smoothly with growing eccentricity as opposed to an abrupt appearance at large $e$.
The next even mode, $m=4$, does appear only at large eccentricity $e \geq 0.5$.
The strength of the odd-modes is consistent with zero for all eccentricities.
This non-existence of odd azimuthal modes, the presence and growing strength of the even modes, and the observed spiral density waves at large $e$, are consistent with the emergence of retrograde resonances in our simulations.
However, these features could also emerge from non-resonant interactions with the time varying potential of the equal mass binary and conclusively disentangling the two would require a more detailed analysis.

%
\subsection{Accretion rate}
\label{s:accretion_rate}

To determine periodicity structure in the accretion rate, we compute a 2D periodogram of the accretion rate time series as measured onto both binary components. To do so we utilized the entire $10^4$ binary orbit accretion rate time series from \texttt{Disco}, spanning from $e=0-0.8$ (similar results are found with \texttt{Sailfish}, see below). As in \citet{Duffell:2020}, we convolve a Gaussian window in time (and so also binary eccentricity) with the inner product of the accretion rate and Fourier vector with angular frequency $\omega$ in frequency space,
\begin{equation}
    \mathcal{P}(e, \omega) =  \frac{1}{\sqrt{2 \pi \sigma^2}} \int^{t(e_f)}_{t(e_0)}{  \mathrm{e}^{-\frac{1}{2}\frac{\left(t(e)-\tau\right)^2}{\sigma^2}}  \Mdot(\tau) \mathrm{e}^{i \omega \tau} d\tau},
    \label{eq:2Dpdgm}
\end{equation}
where eccentricity dependence comes through $t(e)$, the inverse of Eq.~(\ref{Eq:esweep}). We choose $\sigma=30 P_{b}$ and compute the norm of Eq.~(\ref{eq:2Dpdgm}) over a $300 \times 300$ grid of values of $e$ ranging from $0.0$ to $0.8$ and $\omega$ ranging from $ 0.1 \Omega_b$ to $2.5\Omega_b$, corresponding to variability timescales between $(0.1-2.5)P_{\orb}$.

\begin{figure}
 \begin{center}$
    \hspace{-15pt}
    \begin{array}{c}
    \includegraphics[width=\columnwidth]{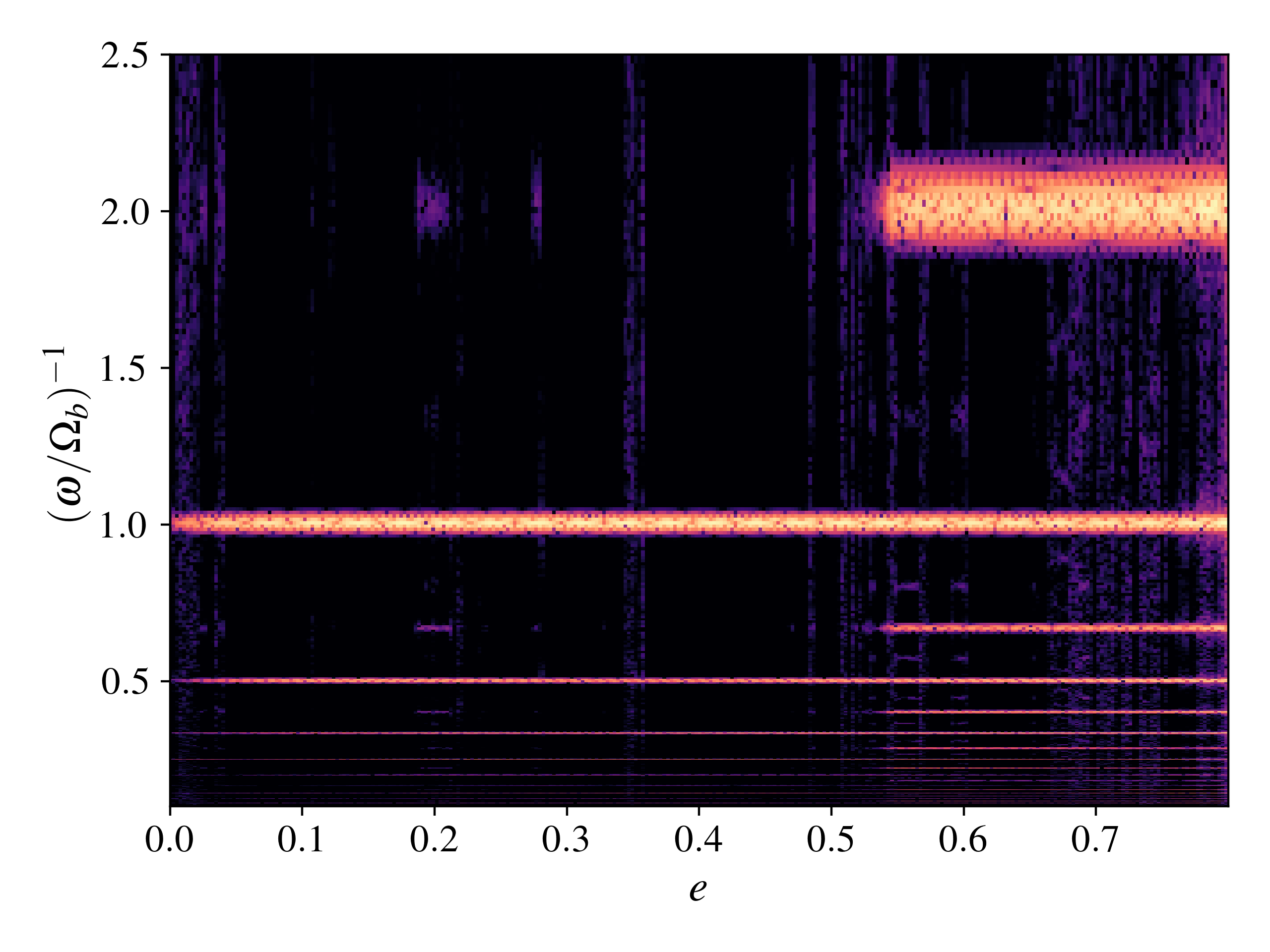}
    \end{array}$
 \end{center}
 \vspace{-20pt}
 \caption{
 Power in the total binary accretion-rate periodogram computed via Equation~\ref{eq:2Dpdgm}. The dominant power at all eccentricities is at the orbital period and its higher frequency harmonics.
 For eccentricities $e \gtrsim 0.55$ power also emerges in a wide band around two-times the orbital period.
 }
 \label{fig:2Dpdgm}
\end{figure}

Figure \ref{fig:2Dpdgm} plots contours of the log-base-10 power computed over this grid. The most prominent power is concentrated in narrow bands at the orbital time and its higher frequency harmonics. In addition, a wider band is concentrated at twice the binary orbital period (half the orbital frequency), appearing for eccentricities above $e\approx0.55$. The latter appearance of a twice-orbital periodicity is notably coincident with the appearance of retrograde resonances as argued in \S~\ref{s:resonances}. The twice-orbital periodicity derives from the alternating process of cavity-clearing described in \S~\ref{s:disk-morphology}.

\begin{figure*}
\begin{center}$
\begin{array}{ccc}
\includegraphics[scale=0.33]{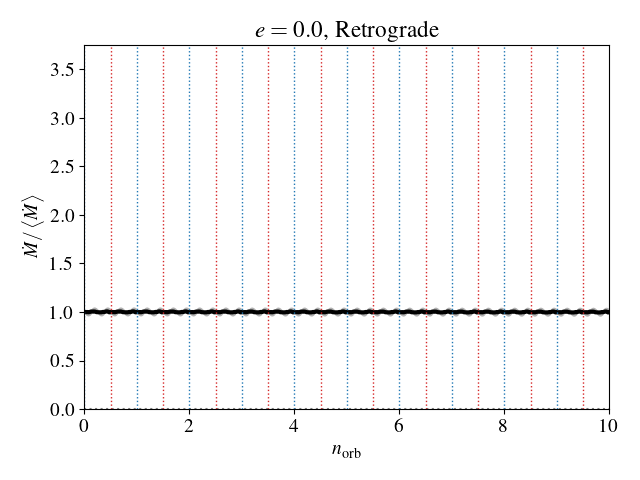} &
\includegraphics[scale=0.33]{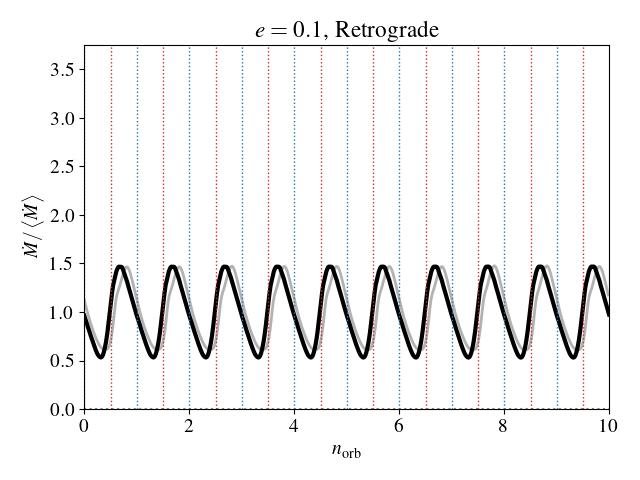} &
\includegraphics[scale=0.33]{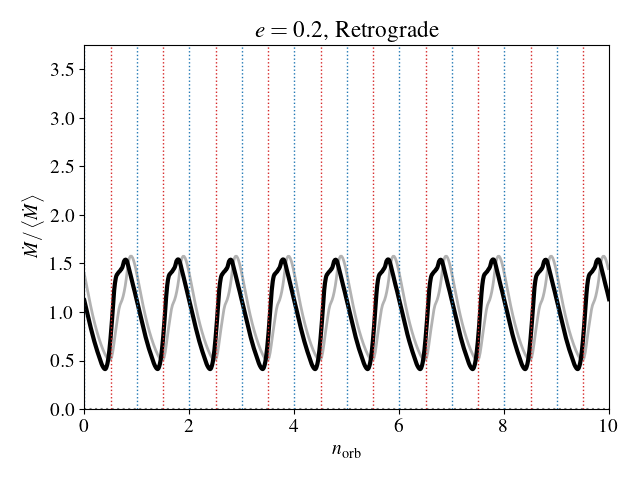} \\
\includegraphics[scale=0.33]{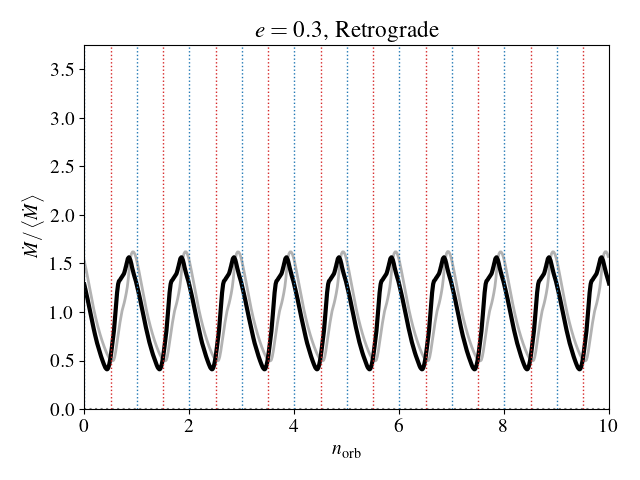} &
\includegraphics[scale=0.33]{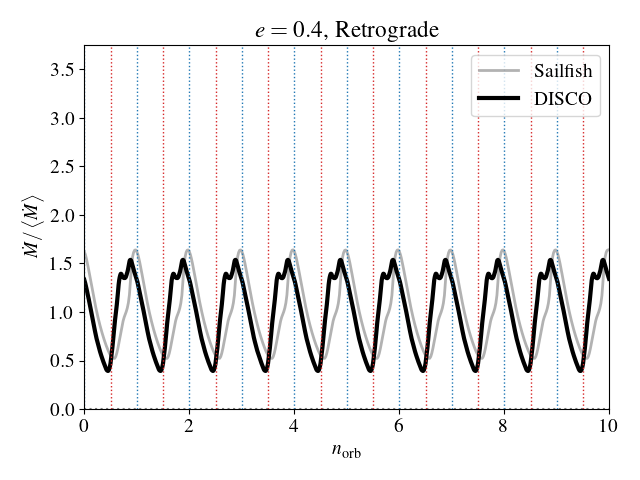} &
\includegraphics[scale=0.33]{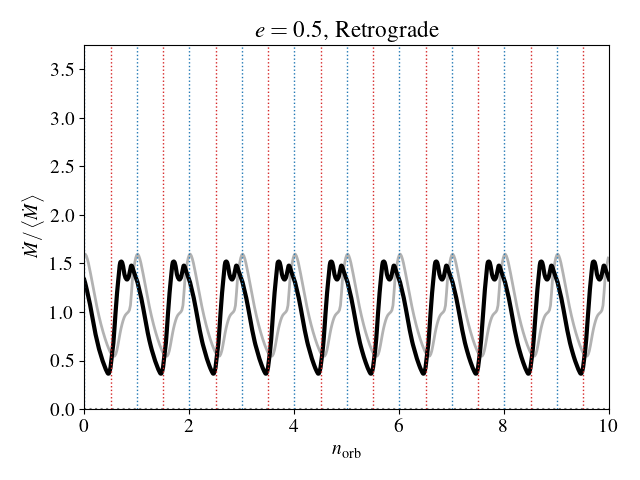} \\
\includegraphics[scale=0.33]{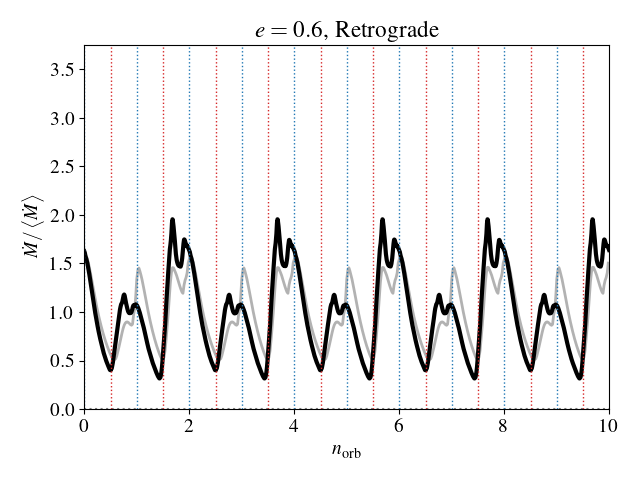} &
\includegraphics[scale=0.33]{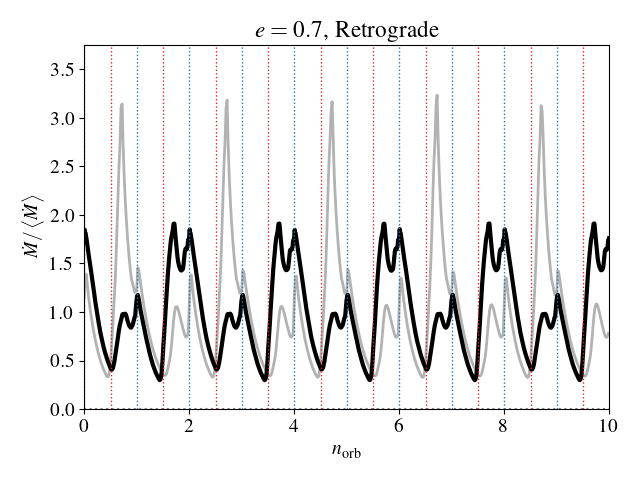} &
\includegraphics[scale=0.33]{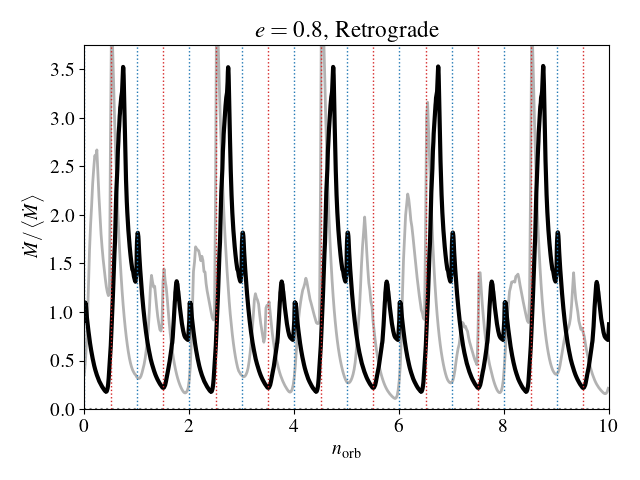} 
\end{array}$
\end{center}
\vspace{-15pt}
\caption{
The total accretion rate measured onto the binary components from the \texttt{Disco} sweep (black) and for comparison, the \texttt{Sailfish} sweep (grey). In each panel, accretion rate time series are shown for 10 orbits starting from where the eccentricity sweep (Equation \ref{Eq:esweep}) reaches the binary orbital eccentricity $e$, indicated in each panel. Vertical dotted blue (red) lines denote pericenter (apocenter). While there are small differences between codes, especially at high eccentricities, a few robust qualitative features persist: For lower eccentricities ($e\lesssim0.55$) the accretion rate times series is modulated at the orbital period, with a double peaked (for \texttt{Disco}) or kinked (for \texttt{Sailfish}) structure at intermediate eccentricities. For $e\gtrsim0.55$, a twice-orbital period modulation arises, as indicated by Figure \ref{fig:2Dpdgm}. For the highest eccentricities probed in this study, a large accretion rate spike is induced following apocenter.
}
\label{fig:Mdots_of_e}
\end{figure*}

In Figure \ref{fig:Mdots_of_e} we further explore the time series accretion rates for representative binary eccentricities. In this figure we plot the \texttt{Disco} results in thick black lines and also the \texttt{Sailfish} results in grey lines for comparison. We first describe the \texttt{Disco} accretion-rate time series.
For $e=0$, the accretion rate is steady. As $e$ increases from zero, the accretion rate becomes strongly modulated at the orbital period with peaks just following apocenter, and with amplitude growing with $e$ until $e\sim0.2$. From $e\sim0.2-0.5$, the amplitude of accretion rate modulations saturates to $\sim 50\%$ of the mean, and the time of peak accretion rate drifts from just after apocenter towards just before pericenter with increasing $e$. At $e\gtrsim0.3$, a second peak in the accretion rate develops, occurring just after apocenter and before the other peak, which, for $0.3\lesssim e \lesssim 0.5$, occurs just before pericenter. For $e\gtrsim0.55$, the twice-per-orbit periodicity appears , and the amplitude of modulations begins to again grow with $e$.
 For $e=0.6$ and $e=0.7$, the twice-orbital-period periodicity manifests as the double peaked accretion rate modulation becoming a factor of $\sim2$ higher every other orbit, but otherwise similar in shape, due again to the alternating cavity structure seen in Figure~\ref{fig:e07_tseries} and described in \S~\ref{s:disk-morphology}. For $e=0.8$, the higher accretion-rate modulation occurring every other orbit is punctuated by a large, factor of $3.5$ accretion rate spike just following apocenter.

The \texttt{Sailfish} accretion-rate time series are similar to the \texttt{Disco} results in periodicity structure and in the magnitude of accretion rate modulations, but exhibit a number of differences. Primarily, the double peaked structure does not manifest until $e\gtrsim0.6$ for the \texttt{Sailfish} runs. Rather, at intermediate eccentricities, what appears as a double peaked structure in the \texttt{Disco} runs, appears instead as a low accretion rate kink in the grey \texttt{Sailfish} time series. At high eccentricities \texttt{Sailfish} also exhibits large spikes following apocenter, but starting at $\sim 0.7$ compared to $e\approx 0.8$ for \texttt{Disco}. We note that the \texttt{Sailfish} accretion rate time series in the final $e=0.8$ panel of Figure~\ref{fig:Mdots_of_e}, is likely exhibiting spurious behavior due to the non-explicit conservation of angular momentum.
Hence, while periodicity timescales and the magnitude of accretion-rate variations are robust between the codes, small differences in the shape of the accretion rate time series such as the observed double peaked structure are not, and caution should be taken in applying these to observable features of accreting binaries (in addition to further complications in conversion of accretion rates to luminosities).

\section{Discussion}
\label{S:Discussion}
%

%
\subsection{Comparison with previous studies} \label{s:study_comp}

The results presented in this study generally agree qualitatively with previous numerical and analytic works on retrograde binary accretion, but there exist pertinent differences.
Analytic modelling of retrograde accretion scenarios from \cite{NixonKingPringle:2011, RoedigSG_retro:2014, Schnittman_Krolik_retro:2015} were all generally founded upon the assumption that binary evolution is dominated by the physical accretion of counter-rotating gas at binary apocenter (and sometimes at pericenter). 
Generally, these models accurately predict that the binary will shrink its semi-major axis and grow its eccentricity at nearly all eccentricities.
However, \cite{NixonKing_retro+2011} forecasted that at small eccentricities $e < h/r$, the binary would actually shrink its eccentricity and remain near circular due to opposing accretion effects at binary apocenter and pericenter.
This study does not corroborate this prediction and finds that the binary grows its eccentricity always.
On the contrary, the modelling of \cite{Schnittman_Krolik_retro:2015} anticipates binary eccentricity growth at all eccentricities, but quantitatively predicts the effect to be largest at small eccentricities and to decline as eccentricity grows.
The results in Figure \ref{fig:adot_edot} show the opposite.
Eccentricity growth is smallest at small $e$ and grows to a peak value at $e\sim 0.15$ after which it stabilizes to $d e / d\log{M} \sim 2$.

\begin{figure*}
 \centering
  \includegraphics{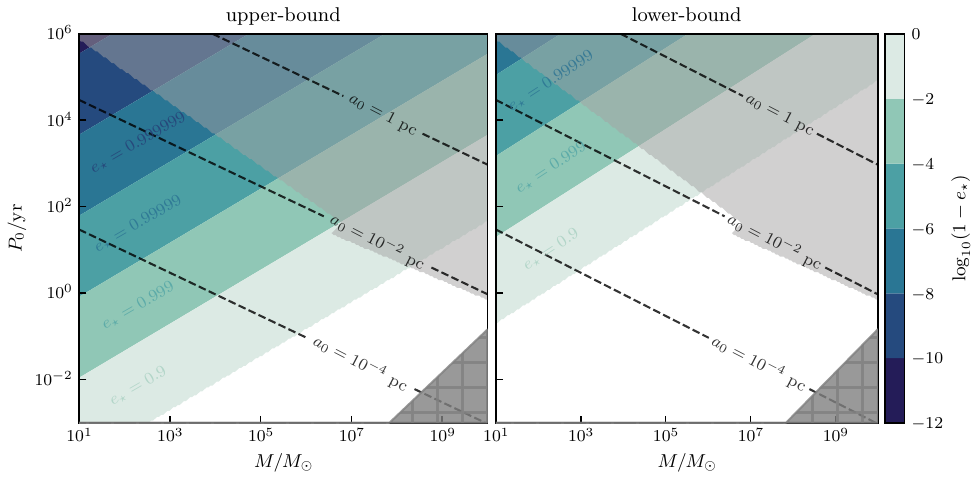}
  \vspace{-10pt}
  \caption{Upper- and lower-bounds (\emph{left} and \emph{right} respectively) to the eccentricity achieved by MBHB's accreting from a retrograde disk determined according to Equation \ref{eq:emax}. The light shaded region in the upper-right of each panel shows the region where a binary of that mass and initial semi-major axis does not fit in a gravitationally stable disk. The dark grey shaded region in the bottom right corner shows where the binary does not fit outside the component ISCO's for a given mass.
  The dashed black lines illustrate initial semi-major axes $a_0$ (in parsec) associated with the initial binary periods $P_0$ shown on the y-axes.}
 \label{fig:emax}
\end{figure*}

These studies (along with \citealt{BankertShiKrolik:2015, NixonLubow:RetroRes:2015}) have additionally either verified or motivated their models with particle- and grid-based numerical studies of accreting retrograde systems.
These simulations generally substantiated the postulate that the orbital evolution was dominated by the physical capture of gas at binary apocenter, and that gravitational forces were a subdominant effect.
We have tested this directly in our full-domain solutions and found that the opposite is true: the binary orbital evolution is almost entirely explained by gravitational forces on the binary, and the physical accretion of mass and momentum is a secondary effect.
Moreover, the dominant phase for driving eccentricity occurs during orbital waxing, between pericenter and apocenter, and the effect at apocenter is small.
Similar to prograde scenarios, the evolution of the binary is controlled by time-dependent, non-axisymmetric features in the disk morphology that exert strong gravitational forces,
and we have demonstrated that the strongest forces come from the intra-orbital material that passes within the semi-major axis of the binary $r < a$ .
We hypothesize that this had previously been missed because older simulations did not include the inner-most regions of the accretion flow either by construction \citep{BankertShiKrolik:2015} or because of the specifics of their numerical scheme and setup \citep{NixonKing_retro+2011, RoedigSG_retro:2014, NixonLubow:RetroRes:2015}.
This is akin to using a sink-prescription that is too large and aggressive (see Appendix) such that it washes away the complex interplay between intra-obit material and the circumbinary flow as well as the non-trivial exchanges of energy and angular momentum between this material and the binary
(as was noted for small mass-ratio binaries $q < 0.1$ by \citealt{Amaro-Seoane_Retro+2016}).

%
\subsection{Observational implications}

The presence of circumbinary disks can leave observable imprints in both electromagnetic and gravitational wave radiation from accreting binaries, and the effects of such accretion phases leave a lasting impact on populations of binaries that have undergone gas-mediated phases.
Such effects have been discussed in detail for prograde accretion scenarios \citep[\eg,][]{Farris:2015:GW, Geoff:2017, Bortolas:2021, Krauth:2023}, and many retrograde effects have been discussed in previous works \citep[c.f.][]{Schnittman_Krolik_retro:2015}.
We expand on these by discussing the implications of this study in the context of recent prograde results.

\emph{Disk-mediated decay and GWs} ---
In prograde solutions, it is possible for accretion-mediated phases of binary evolution to both expand the binary orbit and to facilitate binary inspiral. 
Namely, near-circular-orbit binaries tend to circularize and expand their orbits due to accretion and interaction with a CBD for disk scale-heights $h/r \gtrsim 0.05$; however, prograde binaries with initial eccentricity $e \gtrsim 0.08$ evolve towards an equilibrium eccentricity of $e_{\rm eq} \sim 0.4$ (or some value $0.25 \lesssim e_{\rm eq} \lesssim 0.5$ for $q < 1$ \citealt{Siwek:2023}) where they shrink their semi-major axis at a rate $\mathcal{A} \equiv d\log a/ d\log M$ that is order unity \citep{Zrake+2021, DOrazioDuffell:2021}.

For retrograde accretion, we can describe the orbital evolution with a simple fitting formula,
\begin{eqnarray}
\edot &=& \mathcal{E} \frac{\dot M_0}{M} 
  \begin{cases} 
      e & e < e_* \\
      e_{*} & e \geq e_*
   \end{cases} \nonumber \\
\frac{\adot}{a} &=& \mathcal{A} \frac{\dot M_0}{M}   .
\end{eqnarray} 
For the remaining discussion we choose $\mathcal{E}=30$, $e_*=0.1$, and $\mathcal{A}=-10$. Note, however, that a range of $e_* \sim 0.1-0.2$ and $\mathcal{E}e_*=2.5-3.5$ are consistent with the \texttt{Disco} and \texttt{Sailfish} results in Figure \ref{fig:adot_edot}.
Hence, in contrast to the prograde results, and in accordance with previous retrograde accretion studies, retrograde accretion scenarios facilitate binary decay at all eccentricities, and do so at a rate $\mathcal{A} \approx -10$; a factor of $2-3$ faster than prograde disks at $e_{\rm eq}$.
For black holes accreting at their Eddington rate, this implies a retrograde gas-mediated inspiral timescale $\tau_a = a / \dot a = -\mathcal{A}^{-1} M / \dot M \sim 4.5 \,{\rm Myr}$ where we've used $M / \dot M \approx 4.5 \times 10^{7}\, {\rm yr}$ as the Eddington-limited mass doubling time of an accreting black hole---or Salpeter time---with an accretion efficiency 0.1. This is shorter than the expected $10-100 \,{\rm Myr}$ lifetimes of quasars \citep{PMartini:2004}.

In contrast with prograde solutions, binaries accreting from retrograde disks have their eccentricity pumped at all eccentricities. 
Our $\edot(e)$ solution is nearly linear for $e \lesssim 0.1$, with an estimated form $30e\, \Mdot/M $, and approximately constant for $e\geq0.1$. 
Therefore, for initial eccentricities $e_0\lesssim0.1$, the binary will grow its eccentricity exponentially, with an e-folding timescale of $(\sim 30 \Mdot/M)^{-1}$, quickly driving initially small eccentricities into the constant $\edot$ regime, where eccentricity grows at approximately twice the mass doubling rate. Hence, for any initial binary eccentricity, retrograde circumbinary disks will drive eccentricities $e\rightarrow 1$ in approximately half of a Salpeter time, $\sim 20$ Myr.  
Comparing the mutual evolution of $\dot a$ and $\dot e$, this implies that $a$ will decrease by 5 e-foldings $a_0 \mathrm{e}^{-5}$ in the time required for $e \rightarrow 1$.

A retrograde disk will simultaneously pump binary eccentricity and shrink the semi-major axis until the effects of energy and angular momentum loss by GWs begin to dominate at either high-$e$ or small-$a$. 
In the large eccentricity limit, there exists some eccentricity $e_\star$ at which GWs will begin to damp binary eccentricity at a rate faster than the disk can pump it.
We can estimate a \emph{lower-bound} for the eccentricity achieved from retrograde accretion before GWs take over by finding $e_\star$ such that
\begin{equation}
    \dot e_{GW}(M, a_0\mathrm{e}^{-5}, e_\star) = -\dot e_{CBD}(e_\star)
 \label{eq:emax}
\end{equation}
as a function of binary mass $M$ and initial separation $a_0$ (assuming $q=1$).
This lower bound is likely the more realistic scenario, but we also compute
an \emph{upper-bound} for binary eccentricity at the onset of a GW dominated regime by using $a_0$ instead of $a_0 e^{-5}$.
This ignores the commensurate shrinking of semi-major axis during eccentricity pumping\footnote{We note that the quantities $a(1 + e)$ and $a(1 - e)$ are not conserved and are always decaying.}.
These upper- and lower-bounds on the disk-driven eccentricity are shown as contours of $\log_{10}(1 - e_\star)$ in $a_0-M$ space in Figure \ref{fig:emax} (the \emph{left} and \emph{right} panels respectively).
The light shaded region in the upper right corner denotes where $a_0$ would not fit into a gravitationally stable, steady-state thin disk (using Equation 16 in \citealt{Haiman+2009}), and the dark shaded region in the bottom right corner illustrates where a binary of given mass is within the component ISCOs. The black-dashed lines show initial semi-major axes associated to the initial binary period $P_0$.

Massive binaries that undergo periods of prograde accretion before merging in the LISA band are predicted to be driven to an equilibrium eccentricity $e_{\rm eq} = 0.4$.
Binaries that enter a GW-dominated regime at $e_{\rm eq}$ will retain only a sub-percent eccentricity very near the LISA detection threshold \citep{Cuadra:2009, Zrake+2021}.
Massive binaries that have their eccentricity pumped to near unity by retrograde accretion, however, may retain significantly more eccentricity upon entering the LISA band and are comparatively more likely to lie above the detector's eccentricity detection threshold \citep{MuditGarg:2023:LISA_e_Detect}.
Assuming the lower-bound as the more realistic scenario in Figure \ref{fig:emax}, we still observe that phases of retrograde accretion could drive very large eccentricities $e_\star \gtrsim 0.9$ into massive binaries that are eventually detected by LISA, if accretion commences at large enough separations $a_0 \sim 10^{-2}$pc.
Moreover, because we observe retrograde minidisks, binaries that have undergone retrograde accretion will acquire spins that are counter-aligned with the binary orbital angular momentum.
Thus, a signpost of such retrograde accretion phases may be the measurement of notable eccentricity in conjunction with negative effective spin parameters $\chi_{\rm ef} < 0$ in the GW signal analysis.

Gravitational wave emission at such large eccentricities will also shift GW power from the standard, circular $2f_b$---with $f_b$ the binary frequency---to higher frequencies $f = f_b (1+e)^{1/2} / (1 - e)^{3/2}$.
Rapid environment driven coalescence will additionally diminish GW power at frequencies where such effects are active.
For the most massive binaries $M > 10^8 \, M_\odot$, like those posited to source the low-frequency gravitational wave background \citep{NANOG-GWB:2023, EPTA-GWB:2023, ParkesPTA-GWB:2023, CPTA-GWB:2023}, the effect of a retrograde accretion phase that both pumps eccentricity and rapidly shrinks the binary orbit would be to reduce GW power at the low-frequency end of the background, as was tentatively observed \citep{NANOG-SMBHBs:2023, EPTA-SMBHBs:2023}.
However, Figure \ref{fig:emax} suggests that eccentricity pumping for such massive binaries may not be as effective as for smaller systems, and that self-gravitating disks, where our results break down \citep[c.f.][]{Franchini_SG:2021}, likely become relevant. Further modeling of the effects of retrograde accretion on low-frequency gravitational wave backgrounds is warranted.

It is worth mentioning that varying the disk scale-height $h/r$ can dramatically alter the standard circular prograde picture presented above, as disks with $h/r \lesssim 0.05$ recover binary inspiral \citep{Tiede:2020, Penzlin:2022} and show tentative evidence that those with values approaching the expected theoretical limit of $h/r \sim 0.01-0.001$ for geometrically thin disks around massive black holes possess orders of magnitude faster inspiral rates $\mathcal{A} \approx \mathcal{O}(10^2)$ \citep{Dittmann:2022}.
The dependence of eccentricity evolution for such thin disks, however, is yet to be explored, and it remains unclear how changing the scale-height in retrograde scenarios changes the solutions.

\emph{Electromagnetic observations} --- 
While 2D isothermal hydrodynamics simulations cannot self-consistently produce lightcurves of the accretion flow, the accretion rate can serve as an approximation for the variability in the system luminosity, and recent magnetohydrodynamics simulations have shown that the variability of Poynting fluxes from component jets track the binary accretion rate \citep[\eg][]{Combi:2022}.
One of the most prominent features of prograde accretion solutions are the presence of a 5 orbit accretion variability associated with an $m=1$ density mode that orbits the inner edge of the prograde CBD for near-circular binaries.
By contrast, the accretion rate for circular retrograde binaries shows almost no variability, except for small fluctuations at the orbital frequency.
As documented in \S \ref{s:disk-morphology} and \S \ref{s:resonances}, retrograde solutions also retain a high-degree of axisymmetry for $e < 0.55$, and as such are strongly modulated only on the orbital frequency of the binary; and the periodic variability for eccentricities in this range are almost indistinguishable, with the exception of a possible double peak emerging for $e = 0.4-0.5$, making them hard to distinguish in time-domain observational data.
At large eccentricities $e > 0.55$, however, retrograde binary accretion manifests a two-orbit double peaked accretion signature characterized by a large flare, followed by a flare of approximately half the magnitude.
This kind of behavior has not been observed in any prograde configurations and could serve as a unique observational signature of retrograde binary accretion.
At the highest eccentricities this ``double flaring'' feature could manifest as quasi-periodic (or periodic if temporally and flux resolved) eruptions on timescales of super-massive black hole binary orbits (days to years).

Beyond variability in the accretion rate and system luminosity, the existence of axisymmetric density waves propagating through the CBD for $0 < e \lesssim 0.55$ may also have observational implications.
The higher density rings, with possibly higher temperature when considering non-isothermal equations of state, would cause a time dependent variation of the disk spectra compared to that of a steady disk, causing the opposite of a spectral notch proposed for prograde disks \citep{GultekinMiller_SEDGaps:2012, Roedig_SEDsigs+2014}, but also varying periodically in time. 
In addition, these retrograde ring waves may manifest in images of circumbinary disks around stellar binaries, resolvable with instruments such as the Atacama Large Millimeter Array \citep[ALMA, \eg,][]{Alves+2019}. Synthetic observations of retrograde systems should explore this possibility \citep[\eg,][]{Ragusa+2021}, as recent theoretical work predicts $10\%$ of stellar circumbinary disk systems to form retrograde \citep{Elsender+Bate:2023}.

%
\subsection{Numerical limitations}

In order to densely sample the binary eccentricity and accurately calculate the orbital evolution, we have made a number of simplifying assumptions.
Foremost, these simulations have employed an isothermal treatment of the gas thermodynamics where any heat generated from viscosity or shocks is assumed to instantly leave the system as radiation.
In future work it would be informative to include an equation of state that accounts for such heating processes and subsequently cools the disk on an appropriate timescale.
We have also restricted ourselves to only two dimensional, co-planar configurations, but possible out of plane effects have been previously reported \citep{NixonKingPringle:2011, RoedigSG_retro:2014}.
We have additionally ignored magnetic fields and considered only a simple constant-$\nu$ model for the disk viscosity. A more detailed treatment would include electromagnetic effects and resolve the magneto-rotational turbulence that would self-consistently govern angular momentum transport in the disk.

The other major simplification of this study is that it focused exclusively on equal mass-ratio binaries. Because of the complexity and importance of the inner-most regions of the accretion flow to our results, we posit that varying $q$ may produce substantially different results and plan to explore this in future work.

\section{Conclusions}
\label{S:Conclusions}

We have conducted a number of simulations of equal mass binaries of varying eccentricity accreting from retrograde circumbinary disks using the grid based hydrodynamics codes \texttt{Disco} and \texttt{Sailfish}.
We have continuously characterized the effects of such retrograde accretion on the binary's orbital elements, documented the morphology and phase-dependent behavior of retrograde accretion flows at numerous eccentricities, and determined observational signatures associated with retrograde solutions.

In accordance with most theoretical expectation and previous analytic work, we find that retrograde accretion scenarios shrink the binary semi-major axis at all eccentricities with a near-constant rate $d\log a / d\log M = -10$, and simultaneously pump eccentricity for any $0 < e < 0.8$ (Figure \ref{fig:adot_edot}).
The latter point, however, is in contrast with previous SPH simulations and impulse-approximation based models that predicted near-circular binaries would have their eccentricity damped, remaining effectively circular.
Moreover, we find that the dominant contribution to the binary orbital evolution is gravitational forces from the retrograde binary minidisks and retrograde-bridge that form in the inner-most $r < a$ of the accretion flow (Figures \ref{fig:decomposition} and \ref{fig:torque_power}).
Asymmetries in these features yield gravitational forces that act at all phases of the binary orbit, and the primary contributions to orbital eccentricity pumping occur during binary waxing (Figures \ref{fig:e03_tseries}, \ref{fig:e07_tseries}, and \ref{fig:edot_v_phase}).

We found that the morphology of the retrograde CBD's posses three separate regimes as we vary binary eccentricity (see Figure \ref{fig:sigma_maps}). 
First, near-circular binaries $e < 0.01$ display a predominantly time-invariant structure in the co-rotating frame of the binary, with the exception of small wiggles in the retrograde-bridge (Figure \ref{fig:e00_snapshot}).
In regime (\emph{ii}), binaries with eccentricities $0.025 < e < 0.55$ yield phase-dependent disk behavior that repeats every binary orbit and is characterized by an axisymmetric density wave that is driven into the disk once per orbit during binary waning (Figure \ref{fig:e03_tseries}).
Lastly, binaries with $e > 0.55$ in regime (\emph{iii}) manifest two-orbit periodic disk oscillations characterized by the forcing of non-axisymmetric $m=2$ spiral density waves in the ``first'' orbit and a predominantly axisymmetric density ring (with comparatively weak spiral arms) in the ``second'' orbit (Figure \ref{fig:e07_tseries}).
The emergence of even-azimuthal-mode spiral density waves at $e > 0.55$ is consistent with previous predictions that disk resonances can manifest in viscous, retrograde disks at large eccentricities (see also \S~\ref{s:resonances} and Figure \ref{fig:moments}).

We have additionally analyzed the characteristic variability of the accretion rates for binaries of many different eccentricities in retrograde CBD's (Figure \ref{fig:Mdots_of_e}).
We observed that circular, retrograde solutions are dramatically different from their prograde counterparts as they exhibit almost no accretion variability due to their functionally steady-state configurations.
Binary eccentricities in retrograde regime (\emph{ii}) all exhibit strong accretion rate modulation at the orbital frequency with the possible emergence of a kinked or double peaked structure at $e = 0.4-0.5$.
For eccentricities in regime (\emph{iii}), $e > 0.55$, accretion rate modulations start varying at both the binary period and twice the binary period as the accretion oscillates between minor- and major-spikes.

In light of modern numerical studies that have begun 
characterizing the behavior of prograde circumbinary accretion across system parameters like eccentricity, mass ratio, and disk thickness, we have revisited the complimentary retrograde scenario.
Occurrences of retrograde accretion can be an important contributor to binary orbital decay and eccentricity growth, and they exhibit unique observational features for binary searches in large-scale surveys.
We have demonstrated that high resolution, full-domain treatments are required to accurately quantify these effects and how they contrast with prograde solutions.
As such, retrograde investigations should continue alongside their prograde counterparts as the community continues to characterize these systems.

\section*{Acknowledgements}

The authors sincerely thank Jonathan Zrake and Paul Duffell for making their codes \texttt{Sailfish} and \texttt{Disco} available for use in this work.
C.T. and D.J.D extend their gratitude to all members of the GW-Astro group at the NBIA and Jeff J. Andrews for useful discussions.
We also thank the anonymous referee for a constructive report.
D.J.D. received funding from the European Union's Horizon 2020 research and innovation programme under Marie Sklodowska-Curie grant agreement No. 101029157.  D.J.D. and C.T. received support from the Danish Independent Research Fund through Sapere Aude Starting Grant No. 121587.

\section*{Data Availability}
The data used for this study can be shared upon reasonable request to the corresponding author.
The codes used for the presented simulations are also publicly available on Github at 
\url{https://github.com/duffell/Disco}
and
\url{https://github.com/clemson-cal/sailfish}.

\renewcommand\thefigure{A\arabic{figure}}
\setcounter{figure}{0}
\begin{figure*}
  \includegraphics{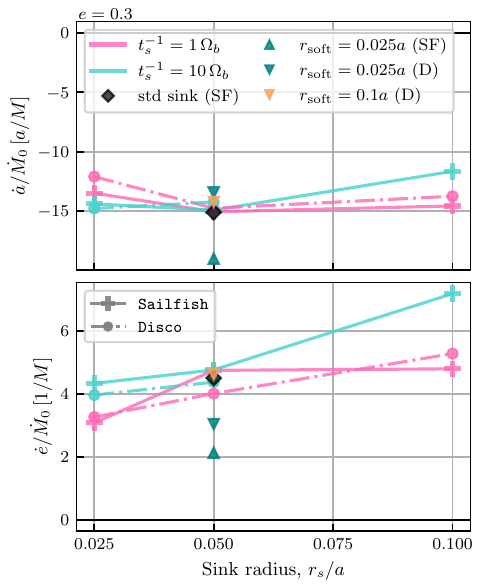}
  \includegraphics{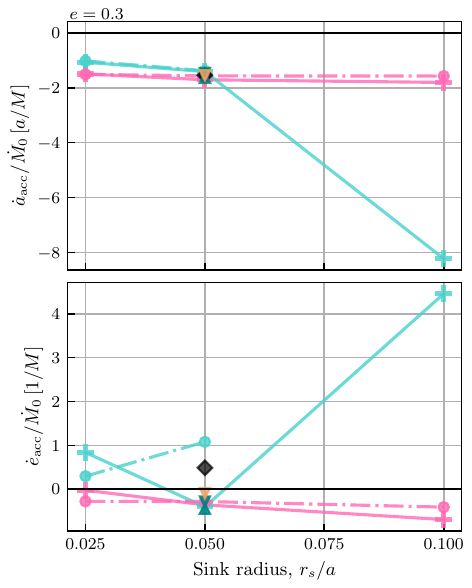}
  \caption{The effects of sink radius, sink rate, and gravitational softening on the full orbital evolution of the binary (\emph{left}) and the component of evolution due to accretion alone (\emph{right}) at fixed binary eccentricity $e=0.3$.
  The pink curve shows the fiducial sink rate at varying sink radius for both \texttt{Disco} (\emph{dashed-dotted + circles}; also marked D in the supplementary runs that vary the softening radius) and \texttt{Sailfish} (\emph{solid + crosses}; SF in the supplementary runs).
  The cyan curve shows the same experiment for a fast-sink.
  The black diamond shows the use of a ``standard'' acceleration-free sink, and the triangles illustrate the effects of varying the gravitational softening radius.
  Note the change in normalization to $\dot M_0$ vs. the measured $\dot M$. We observe that our results remain largely insensitive to the sink radius and rate so long as the sink region is sufficiently resolved; with the exception of large-fast sinks.}
 \label{fig:sink-test}
\end{figure*}

\bibliographystyle{mnras} 
\bibliography{refs}

\begin{thebibliography}{}
\makeatletter
\relax
\def\mn@urlcharsother{\let\do\@makeother \do\$\do\&\do\#\do\^\do\_\do\%\do\~}
\def\mn@doi{\begingroup\mn@urlcharsother \@ifnextchar [ {\mn@doi@}
  {\mn@doi@[]}}
\def\mn@doi@[#1]#2{\def\@tempa{#1}\ifx\@tempa\@empty \href
  {http://dx.doi.org/#2} {doi:#2}\else \href {http://dx.doi.org/#2} {#1}\fi
  \endgroup}
\def\mn@eprint#1#2{\mn@eprint@#1:#2::\@nil}
\def\mn@eprint@arXiv#1{\href {http://arxiv.org/abs/#1} {{\tt arXiv:#1}}}
\def\mn@eprint@dblp#1{\href {http://dblp.uni-trier.de/rec/bibtex/#1.xml}
  {dblp:#1}}
\def\mn@eprint@#1:#2:#3:#4\@nil{\def\@tempa {#1}\def\@tempb {#2}\def\@tempc
  {#3}\ifx \@tempc \@empty \let \@tempc \@tempb \let \@tempb \@tempa \fi \ifx
  \@tempb \@empty \def\@tempb {arXiv}\fi \@ifundefined
  {mn@eprint@\@tempb}{\@tempb:\@tempc}{\expandafter \expandafter \csname
  mn@eprint@\@tempb\endcsname \expandafter{\@tempc}}}

\bibitem[\protect\citeauthoryear{{Agazie} et~al.,}{{Agazie}
  et~al.}{2023a}]{NANOG-SMBHBs:2023}
{Agazie} G.,  et~al., 2023a, \mn@doi [arXiv e-prints]
  {10.48550/arXiv.2306.16220}, \href
  {https://ui.adsabs.harvard.edu/abs/2023arXiv230616220A} {p. arXiv:2306.16220}

\bibitem[\protect\citeauthoryear{{Agazie} et~al.,}{{Agazie}
  et~al.}{2023b}]{NANOG-GWB:2023}
{Agazie} G.,  et~al., 2023b, \mn@doi [\apjl] {10.3847/2041-8213/acdac6}, \href
  {https://ui.adsabs.harvard.edu/abs/2023ApJ...951L...8A} {951, L8}

\bibitem[\protect\citeauthoryear{{Alves}, {Caselli}, {Girart}, {Segura-Cox},
  {Franco}, {Schmiedeke}  \& {Zhao}}{{Alves} et~al.}{2019}]{Alves+2019}
{Alves} F.~O.,  {Caselli} P.,  {Girart} J.~M.,  {Segura-Cox} D.,  {Franco}
  G.~A.~P.,  {Schmiedeke} A.,   {Zhao} B.,  2019, \mn@doi [Science]
  {10.1126/science.aaw3491}, \href
  {https://ui.adsabs.harvard.edu/abs/2019Sci...366...90A} {366, 90}

\bibitem[\protect\citeauthoryear{{Amaro-Seoane}, {Maureira-Fredes}, {Dotti}  \&
  {Colpi}}{{Amaro-Seoane} et~al.}{2016}]{Amaro-Seoane_Retro+2016}
{Amaro-Seoane} P.,  {Maureira-Fredes} C.,  {Dotti} M.,   {Colpi} M.,  2016,
  \mn@doi [\aap] {10.1051/0004-6361/201526172}, \href
  {https://ui.adsabs.harvard.edu/abs/2016A&A...591A.114A} {591, A114}

\bibitem[\protect\citeauthoryear{{Antoniadis} et~al.,}{{Antoniadis}
  et~al.}{2023a}]{EPTA-GWB:2023}
{Antoniadis} J.,  et~al., 2023a, \mn@doi [arXiv e-prints]
  {10.48550/arXiv.2306.16214}, \href
  {https://ui.adsabs.harvard.edu/abs/2023arXiv230616214A} {p. arXiv:2306.16214}

\bibitem[\protect\citeauthoryear{{Antoniadis} et~al.,}{{Antoniadis}
  et~al.}{2023b}]{EPTA-SMBHBs:2023}
{Antoniadis} J.,  et~al., 2023b, \mn@doi [arXiv e-prints]
  {10.48550/arXiv.2306.16227}, \href
  {https://ui.adsabs.harvard.edu/abs/2023arXiv230616227A} {p. arXiv:2306.16227}

\bibitem[\protect\citeauthoryear{Bankert, Krolik  \& Shi}{Bankert
  et~al.}{2015}]{BankertShiKrolik:2015}
Bankert J.,  Krolik J.~H.,   Shi J.,  2015, The Astrophysical Journal, 801, 114

\bibitem[\protect\citeauthoryear{Barnes \& Hernquist}{Barnes \&
  Hernquist}{1996}]{Barnes:1996}
Barnes J.~E.,  Hernquist L.,  1996, \mn@doi [\apj] {10.1086/177957}, 471, 115

\bibitem[\protect\citeauthoryear{{Bate}, {Lodato}  \& {Pringle}}{{Bate}
  et~al.}{2010}]{BateLodatoPringle:2010}
{Bate} M.~R.,  {Lodato} G.,   {Pringle} J.~E.,  2010, \mn@doi [\mnras]
  {10.1111/j.1365-2966.2009.15773.x}, \href
  {https://ui.adsabs.harvard.edu/abs/2010MNRAS.401.1505B} {401, 1505}

\bibitem[\protect\citeauthoryear{{Bortolas}, {Franchini}, {Bonetti}  \&
  {Sesana}}{{Bortolas} et~al.}{2021}]{Bortolas:2021}
{Bortolas} E.,  {Franchini} A.,  {Bonetti} M.,   {Sesana} A.,  2021, \mn@doi
  [\apjl] {10.3847/2041-8213/ac1c0c}, \href
  {https://ui.adsabs.harvard.edu/abs/2021ApJ...918L..15B} {918, L15}

\bibitem[\protect\citeauthoryear{{Combi}, {Lopez Armengol}, {Campanelli},
  {Noble}, {Avara}, {Krolik}  \& {Bowen}}{{Combi} et~al.}{2022}]{Combi:2022}
{Combi} L.,  {Lopez Armengol} F.~G.,  {Campanelli} M.,  {Noble} S.~C.,  {Avara}
  M.,  {Krolik} J.~H.,   {Bowen} D.,  2022, \mn@doi [\apj]
  {10.3847/1538-4357/ac532a}, \href
  {https://ui.adsabs.harvard.edu/abs/2022ApJ...928..187C} {928, 187}

\bibitem[\protect\citeauthoryear{Cuadra, Armitage, Alexander  \&
  Begelman}{Cuadra et~al.}{2009}]{Cuadra:2009}
Cuadra J.,  Armitage P.~J.,  Alexander R.~D.,   Begelman M.~C.,  2009, \mn@doi
  [\mnras] {10.1111/j.1365-2966.2008.14147.x}, 393, 1423

\bibitem[\protect\citeauthoryear{{D'Orazio} \& {Duffell}}{{D'Orazio} \&
  {Duffell}}{2021}]{DOrazioDuffell:2021}
{D'Orazio} D.~J.,  {Duffell} P.~C.,  2021, \mn@doi [\apjl]
  {10.3847/2041-8213/ac0621}, \href
  {https://ui.adsabs.harvard.edu/abs/2021ApJ...914L..21D} {914, L21}

\bibitem[\protect\citeauthoryear{{D'Orazio}, {Haiman}  \&
  {MacFadyen}}{{D'Orazio} et~al.}{2013}]{DHM:2013:MNRAS}
{D'Orazio} D.~J.,  {Haiman} Z.,   {MacFadyen} A.,  2013, \mn@doi [\mnras]
  {10.1093/mnras/stt1787}, 436, 2997

\bibitem[\protect\citeauthoryear{{Dempsey}, {Mu{\~n}oz}  \&
  {Lithwick}}{{Dempsey} et~al.}{2020}]{Dempsey-TFsinks:2020}
{Dempsey} A.~M.,  {Mu{\~n}oz} D.,   {Lithwick} Y.,  2020, \mn@doi [\apjl]
  {10.3847/2041-8213/ab800e}, \href
  {https://ui.adsabs.harvard.edu/abs/2020ApJ...892L..29D} {892, L29}

\bibitem[\protect\citeauthoryear{{Dittmann} \& {Ryan}}{{Dittmann} \&
  {Ryan}}{2021}]{Dittmann:2021}
{Dittmann} A.~J.,  {Ryan} G.,  2021, \mn@doi [\apj] {10.3847/1538-4357/ac1bbd},
  \href {https://ui.adsabs.harvard.edu/abs/2021ApJ...921...71D} {921, 71}

\bibitem[\protect\citeauthoryear{{Dittmann} \& {Ryan}}{{Dittmann} \&
  {Ryan}}{2022}]{Dittmann:2022}
{Dittmann} A.~J.,  {Ryan} G.,  2022, \mn@doi [\mnras] {10.1093/mnras/stac935},
  \href {https://ui.adsabs.harvard.edu/abs/2022MNRAS.513.6158D} {513, 6158}

\bibitem[\protect\citeauthoryear{Duffel et~al.,}{Duffel
  et~al.}{2023}]{Duffell:code-comparison}
Duffel P.,  et~al., 2023, {In Prep}

\bibitem[\protect\citeauthoryear{{Duffell}}{{Duffell}}{2016}]{DuffellMHDDISCO:2016}
{Duffell} P.~C.,  2016, preprint, \href
  {http://adsabs.harvard.edu/abs/2016arXiv160503577D} {} (\mn@eprint {arXiv}
  {1605.03577})

\bibitem[\protect\citeauthoryear{{Duffell}, {D'Orazio}, {Derdzinski}, {Haiman},
  {MacFadyen}, {Rosen}  \& {Zrake}}{{Duffell} et~al.}{2020}]{Duffell:2020}
{Duffell} P.~C.,  {D'Orazio} D.,  {Derdzinski} A.,  {Haiman} Z.,  {MacFadyen}
  A.,  {Rosen} A.~L.,   {Zrake} J.,  2020, \mn@doi [\apj]
  {10.3847/1538-4357/abab95}, \href
  {https://ui.adsabs.harvard.edu/abs/2020ApJ...901...25D} {901, 25}

\bibitem[\protect\citeauthoryear{{Dunhill}, {Cuadra}  \& {Dougados}}{{Dunhill}
  et~al.}{2015}]{Dunhill+2015}
{Dunhill} A.~C.,  {Cuadra} J.,   {Dougados} C.,  2015, \mn@doi [\mnras]
  {10.1093/mnras/stv284}, \href
  {http://adsabs.harvard.edu/abs/2015MNRAS.448.3545D} {448, 3545}

\bibitem[\protect\citeauthoryear{{Elsender}, {Bate}, {Lakeland}, {Jensen}  \&
  {Lubow}}{{Elsender} et~al.}{2023}]{Elsender+Bate:2023}
{Elsender} D.,  {Bate} M.~R.,  {Lakeland} B.~S.,  {Jensen} E. L.~N.,   {Lubow}
  S.~H.,  2023, \mn@doi [\mnras] {10.1093/mnras/stad1695}, \href
  {https://ui.adsabs.harvard.edu/abs/2023MNRAS.523.4353E} {523, 4353}

\bibitem[\protect\citeauthoryear{{Farris}, {Duffell}, {MacFadyen}  \&
  {Haiman}}{{Farris} et~al.}{2014}]{Farris:2014}
{Farris} B.~D.,  {Duffell} P.,  {MacFadyen} A.~I.,   {Haiman} Z.,  2014,
  \mn@doi [\apj] {10.1088/0004-637X/783/2/134}, \href
  {http://adsabs.harvard.edu/abs/2014ApJ...783..134F} {783, 134}

\bibitem[\protect\citeauthoryear{{Farris}, {Duffell}, {MacFadyen}  \&
  {Haiman}}{{Farris} et~al.}{2015}]{Farris:2015:GW}
{Farris} B.~D.,  {Duffell} P.,  {MacFadyen} A.~I.,   {Haiman} Z.,  2015,
  \mn@doi [\mnras] {10.1093/mnrasl/slu184}, \href
  {http://adsabs.harvard.edu/abs/2015MNRAS.447L..80F} {447, L80}

\bibitem[\protect\citeauthoryear{{Franchini}, {Sesana}  \& {Dotti}}{{Franchini}
  et~al.}{2021}]{Franchini_SG:2021}
{Franchini} A.,  {Sesana} A.,   {Dotti} M.,  2021, \mn@doi [\mnras]
  {10.1093/mnras/stab2234}, \href
  {https://ui.adsabs.harvard.edu/abs/2021MNRAS.507.1458F} {507, 1458}

\bibitem[\protect\citeauthoryear{{Franchini}, {Lupi}  \& {Sesana}}{{Franchini}
  et~al.}{2022}]{Franchini:2022}
{Franchini} A.,  {Lupi} A.,   {Sesana} A.,  2022, \mn@doi [\apjl]
  {10.3847/2041-8213/ac63a2}, \href
  {https://ui.adsabs.harvard.edu/abs/2022ApJ...929L..13F} {929, L13}

\bibitem[\protect\citeauthoryear{{Garg}, {Tiwari}, {Derdzinski}, {Baker},
  {Marsat}  \& {Mayer}}{{Garg} et~al.}{2023}]{MuditGarg:2023:LISA_e_Detect}
{Garg} M.,  {Tiwari} S.,  {Derdzinski} A.,  {Baker} J.~G.,  {Marsat} S.,
  {Mayer} L.,  2023, \mn@doi [\mnras] {10.1093/mnras/stad3477}, \href
  {https://ui.adsabs.harvard.edu/abs/2023MNRAS.tmp.3335G} {}

\bibitem[\protect\citeauthoryear{Goldreich \& Tremaine}{Goldreich \&
  Tremaine}{1979}]{GT79}
Goldreich P.,  Tremaine S.,  1979, \mn@doi [\apj] {10.1086/157448}, 233, 857

\bibitem[\protect\citeauthoryear{Goldreich \& Tremaine}{Goldreich \&
  Tremaine}{1980}]{GT80}
Goldreich P.,  Tremaine S.,  1980, \mn@doi [\apj] {10.1086/158356}, 241, 425

\bibitem[\protect\citeauthoryear{{G{\"u}ltekin} \& {Miller}}{{G{\"u}ltekin} \&
  {Miller}}{2012}]{GultekinMiller_SEDGaps:2012}
{G{\"u}ltekin} K.,  {Miller} J.~M.,  2012, \mn@doi [\apj]
  {10.1088/0004-637X/761/2/90}, \href
  {https://ui.adsabs.harvard.edu/abs/2012ApJ...761...90G} {761, 90}

\bibitem[\protect\citeauthoryear{{Haiman}, {Kocsis}, {Menou}, {Lippai}  \&
  {Frei}}{{Haiman} et~al.}{2009}]{Haiman+2009}
{Haiman} Z.,  {Kocsis} B.,  {Menou} K.,  {Lippai} Z.,   {Frei} Z.,  2009,
  \mn@doi [Classical and Quantum Gravity] {10.1088/0264-9381/26/9/094032}, 26,
  094032

\bibitem[\protect\citeauthoryear{Hayasaki, Mineshige  \& Sudou}{Hayasaki
  et~al.}{2007}]{Hayasaki:2007}
Hayasaki K.,  Mineshige S.,   Sudou H.,  2007, Publications of the Astronomical
  Society of Japan, 59, 427

\bibitem[\protect\citeauthoryear{{Heath} \& {Nixon}}{{Heath} \&
  {Nixon}}{2020}]{HeathNixon:2020}
{Heath} R.~M.,  {Nixon} C.~J.,  2020, \mn@doi [\aap]
  {10.1051/0004-6361/202038548}, \href
  {https://ui.adsabs.harvard.edu/abs/2020A&A...641A..64H} {641, A64}

\bibitem[\protect\citeauthoryear{{Hobbs}, {Nayakshin}, {Power}  \&
  {King}}{{Hobbs} et~al.}{2011}]{HobbsKing:2011}
{Hobbs} A.,  {Nayakshin} S.,  {Power} C.,   {King} A.,  2011, \mn@doi [\mnras]
  {10.1111/j.1365-2966.2011.18333.x}, \href
  {https://ui.adsabs.harvard.edu/abs/2011MNRAS.413.2633H} {413, 2633}

\bibitem[\protect\citeauthoryear{{King} \& {Pringle}}{{King} \&
  {Pringle}}{2006}]{KingPringle:2006}
{King} A.~R.,  {Pringle} J.~E.,  2006, \mn@doi [\mnras]
  {10.1111/j.1745-3933.2006.00249.x}, \href
  {https://ui.adsabs.harvard.edu/abs/2006MNRAS.373L..90K} {373, L90}

\bibitem[\protect\citeauthoryear{{Kley} \& {Nelson}}{{Kley} \&
  {Nelson}}{2012}]{KleyNelson:2012:rev}
{Kley} W.,  {Nelson} R.~P.,  2012, \mn@doi [\araa]
  {10.1146/annurev-astro-081811-125523}, \href
  {http://adsabs.harvard.edu/abs/2012ARA%26A..50..211K} {50, 211}

\bibitem[\protect\citeauthoryear{{Lai} \& {Mu{\~n}oz}}{{Lai} \&
  {Mu{\~n}oz}}{2022}]{LaiMunoz:Review:2022}
{Lai} D.,  {Mu{\~n}oz} D.~J.,  2022, arXiv e-prints, \href
  {https://ui.adsabs.harvard.edu/abs/2022arXiv221100028L} {p. arXiv:2211.00028}

\bibitem[\protect\citeauthoryear{{Lodato}, {Nayakshin}, {King}  \&
  {Pringle}}{{Lodato} et~al.}{2009}]{Lodato:2009}
{Lodato} G.,  {Nayakshin} S.,  {King} A.~R.,   {Pringle} J.~E.,  2009, \mn@doi
  [\mnras] {10.1111/j.1365-2966.2009.15179.x}, 398, 1392

\bibitem[\protect\citeauthoryear{MacFadyen \& Milosavljevi{\'c}}{MacFadyen \&
  Milosavljevi{\'c}}{2008}]{MacFadyen:2008}
MacFadyen A.~I.,  Milosavljevi{\'c} M.,  2008, \mn@doi [\apj] {10.1086/523869},
  672, 83

\bibitem[\protect\citeauthoryear{{Mahesh}, {McWilliams}  \& {Pirog}}{{Mahesh}
  et~al.}{2023}]{MaheshMcW+2023}
{Mahesh} S.,  {McWilliams} S.~T.,   {Pirog} M.,  2023, \mn@doi [arXiv e-prints]
  {10.48550/arXiv.2305.01533}, \href
  {https://ui.adsabs.harvard.edu/abs/2023arXiv230501533M} {p. arXiv:2305.01533}

\bibitem[\protect\citeauthoryear{{Major Krauth}, {Davelaar}, {Haiman},
  {Westernacher-Schneider}, {Zrake}  \& {MacFadyen}}{{Major Krauth}
  et~al.}{2023}]{Krauth:2023}
{Major Krauth} L.,  {Davelaar} J.,  {Haiman} Z.,  {Westernacher-Schneider}
  J.~R.,  {Zrake} J.,   {MacFadyen} A.,  2023, \mn@doi [arXiv e-prints]
  {10.48550/arXiv.2304.02575}, \href
  {https://ui.adsabs.harvard.edu/abs/2023arXiv230402575M} {p. arXiv:2304.02575}

\bibitem[\protect\citeauthoryear{{Martini}}{{Martini}}{2004}]{PMartini:2004}
{Martini} P.,  2004, Coevolution of Black Holes and Galaxies, \href
  {http://adsabs.harvard.edu/abs/2004cbhg.symp..169M} {p.~169}

\bibitem[\protect\citeauthoryear{{Mayer}}{{Mayer}}{2013}]{Mayer:2013:MBHBGasRev}
{Mayer} L.,  2013, \mn@doi [Classical and Quantum Gravity]
  {10.1088/0264-9381/30/24/244008}, \href
  {http://adsabs.harvard.edu/abs/2013CQGra..30x4008M} {30, 244008}

\bibitem[\protect\citeauthoryear{Milosavljevi{\'c} \&
  Merritt}{Milosavljevi{\'c} \& Merritt}{2003}]{Milosavljevic:2003:EvoMBHB}
Milosavljevi{\'c} M.,  Merritt D.,  2003, \mn@doi [\apj] {10.1086/378086}, 596,
  860

\bibitem[\protect\citeauthoryear{{Miranda}, {Mu{\~n}oz}  \& {Lai}}{{Miranda}
  et~al.}{2017}]{MirandaLai+2017}
{Miranda} R.,  {Mu{\~n}oz} D.~J.,   {Lai} D.,  2017, \mn@doi [\mnras]
  {10.1093/mnras/stw3189}, \href
  {http://adsabs.harvard.edu/abs/2017MNRAS.466.1170M} {466, 1170}

\bibitem[\protect\citeauthoryear{Moody, Shi  \& Stone}{Moody
  et~al.}{2019}]{Moody:2019}
Moody M. S.~L.,  Shi J.-M.,   Stone J.~M.,  2019, \mn@doi [The Astrophysical
  Journal] {10.3847/1538-4357/ab09ee}, 875, 66

\bibitem[\protect\citeauthoryear{{Mu{\~n}oz} \& {Lai}}{{Mu{\~n}oz} \&
  {Lai}}{2016}]{MunozLai:2016}
{Mu{\~n}oz} D.~J.,  {Lai} D.,  2016, \mn@doi [\apj]
  {10.3847/0004-637X/827/1/43}, \href
  {https://ui.adsabs.harvard.edu/abs/2016ApJ...827...43M} {827, 43}

\bibitem[\protect\citeauthoryear{{Mu{\~n}oz}, {Miranda}  \& {Lai}}{{Mu{\~n}oz}
  et~al.}{2019}]{Munoz:2019}
{Mu{\~n}oz} D.~J.,  {Miranda} R.,   {Lai} D.,  2019, \mn@doi [\apj]
  {10.3847/1538-4357/aaf867}, \href
  {http://adsabs.harvard.edu/abs/2019ApJ...871...84M} {871, 84}

\bibitem[\protect\citeauthoryear{Nixon \& Lubow}{Nixon \&
  Lubow}{2015}]{NixonLubow:RetroRes:2015}
Nixon C.,  Lubow S.~H.,  2015, Monthly Notices of the Royal Astronomical
  Society, 448, 3472

\bibitem[\protect\citeauthoryear{{Nixon}, {Cossins}, {King}  \&
  {Pringle}}{{Nixon} et~al.}{2011a}]{NixonKing_retro+2011}
{Nixon} C.~J.,  {Cossins} P.~J.,  {King} A.~R.,   {Pringle} J.~E.,  2011a,
  \mn@doi [\mnras] {10.1111/j.1365-2966.2010.17952.x}, \href
  {https://ui.adsabs.harvard.edu/abs/2011MNRAS.412.1591N} {412, 1591}

\bibitem[\protect\citeauthoryear{Nixon, King  \& Pringle}{Nixon
  et~al.}{2011b}]{NixonKingPringle:2011}
Nixon C.~J.,  King A.~R.,   Pringle J.~E.,  2011b, Monthly Notices of the Royal
  Astronomical Society: Letters, 417, L66

\bibitem[\protect\citeauthoryear{{Orosz}, , {Welsh}  \& {et. al}}{{Orosz}
  et~al.}{2012}]{Orosz:2012Sci}
{Orosz} J.~A.,   {Welsh} W.~F.,   {et. al} 2012, \mn@doi [Science]
  {10.1126/science.1228380}, \href
  {http://adsabs.harvard.edu/abs/2012Sci...337.1511O} {337, 1511}

\bibitem[\protect\citeauthoryear{{Penzlin}, {Kley}, {Audiffren}  \&
  {Sch{\"a}fer}}{{Penzlin} et~al.}{2022}]{Penzlin:2022}
{Penzlin} A. B.~T.,  {Kley} W.,  {Audiffren} H.,   {Sch{\"a}fer} C.~M.,  2022,
  \mn@doi [\aap] {10.1051/0004-6361/202141399}, \href
  {https://ui.adsabs.harvard.edu/abs/2022A&A...660A.101P} {660, A101}

\bibitem[\protect\citeauthoryear{{Pringle}}{{Pringle}}{1991}]{Pringle:1991}
{Pringle} J.~E.,  1991, \mnras, 248, 754

\bibitem[\protect\citeauthoryear{{Ragusa} et~al.,}{{Ragusa}
  et~al.}{2021}]{Ragusa+2021}
{Ragusa} E.,  et~al., 2021, \mn@doi [\mnras] {10.1093/mnras/stab2179}, \href
  {https://ui.adsabs.harvard.edu/abs/2021MNRAS.507.1157R} {507, 1157}

\bibitem[\protect\citeauthoryear{{Reardon} et~al.,}{{Reardon}
  et~al.}{2023}]{ParkesPTA-GWB:2023}
{Reardon} D.~J.,  et~al., 2023, \mn@doi [\apjl] {10.3847/2041-8213/acdd02},
  \href {https://ui.adsabs.harvard.edu/abs/2023ApJ...951L...6R} {951, L6}

\bibitem[\protect\citeauthoryear{{Roedig} \& {Sesana}}{{Roedig} \&
  {Sesana}}{2014}]{RoedigSG_retro:2014}
{Roedig} C.,  {Sesana} A.,  2014, \mn@doi [\mnras] {10.1093/mnras/stu194},
  \href {https://ui.adsabs.harvard.edu/abs/2014MNRAS.439.3476R} {439, 3476}

\bibitem[\protect\citeauthoryear{{Roedig}, {Krolik}  \& {Miller}}{{Roedig}
  et~al.}{2014}]{Roedig_SEDsigs+2014}
{Roedig} C.,  {Krolik} J.~H.,   {Miller} M.~C.,  2014, \mn@doi [\apj]
  {10.1088/0004-637X/785/2/115}, \href
  {https://ui.adsabs.harvard.edu/abs/2014ApJ...785..115R} {785, 115}

\bibitem[\protect\citeauthoryear{{Ryan} \& {MacFadyen}}{{Ryan} \&
  {MacFadyen}}{2017}]{Geoff:2017}
{Ryan} G.,  {MacFadyen} A.,  2017, \mn@doi [\apj]
  {10.3847/1538-4357/835/2/199}, \href
  {https://ui.adsabs.harvard.edu/abs/2017ApJ...835..199R} {835, 199}

\bibitem[\protect\citeauthoryear{{Schnittman} \& {Krolik}}{{Schnittman} \&
  {Krolik}}{2015}]{Schnittman_Krolik_retro:2015}
{Schnittman} J.~D.,  {Krolik} J.~H.,  2015, \mn@doi [\apj]
  {10.1088/0004-637X/806/1/88}, \href
  {https://ui.adsabs.harvard.edu/abs/2015ApJ...806...88S} {806, 88}

\bibitem[\protect\citeauthoryear{{Shakura} \& {Sunyaev}}{{Shakura} \&
  {Sunyaev}}{1973}]{SS73}
{Shakura} N.~I.,  {Sunyaev} R.~A.,  1973, \aap, \href
  {http://adsabs.harvard.edu/abs/1973A%26A....24..337S} {24, 337}

\bibitem[\protect\citeauthoryear{{Shi} \& {Krolik}}{{Shi} \&
  {Krolik}}{2015}]{ShiKrolik:2015}
{Shi} J.-M.,  {Krolik} J.~H.,  2015, \mn@doi [\apj]
  {10.1088/0004-637X/807/2/131}, \href
  {http://adsabs.harvard.edu/abs/2015ApJ...807..131S} {807, 131}

\bibitem[\protect\citeauthoryear{Shi, Krolik, Lubow  \& Hawley}{Shi
  et~al.}{2012}]{ShiKrolik:2012}
Shi J.-M.,  Krolik J.~H.,  Lubow S.~H.,   Hawley J.~F.,  2012, The
  Astrophysical Journal, 749, 118

\bibitem[\protect\citeauthoryear{{Siwek}, {Weinberger}, {Mu{\~n}oz}  \&
  {Hernquist}}{{Siwek} et~al.}{2023a}]{Siwek:2022}
{Siwek} M.,  {Weinberger} R.,  {Mu{\~n}oz} D.~J.,   {Hernquist} L.,  2023a,
  \mn@doi [\mnras] {10.1093/mnras/stac3263}, \href
  {https://ui.adsabs.harvard.edu/abs/2023MNRAS.518.5059S} {518, 5059}

\bibitem[\protect\citeauthoryear{{Siwek}, {Weinberger}  \& {Hernquist}}{{Siwek}
  et~al.}{2023b}]{Siwek:2023}
{Siwek} M.,  {Weinberger} R.,   {Hernquist} L.,  2023b, \mn@doi [\mnras]
  {10.1093/mnras/stad1131}, \href
  {https://ui.adsabs.harvard.edu/abs/2023MNRAS.522.2707S} {522, 2707}

\bibitem[\protect\citeauthoryear{{Sudarshan}, {Penzlin}, {Ziampras}, {Kley}  \&
  {Nelson}}{{Sudarshan} et~al.}{2022}]{Sudarshan:2022}
{Sudarshan} P.,  {Penzlin} A. B.~T.,  {Ziampras} A.,  {Kley} W.,   {Nelson}
  R.~P.,  2022, \mn@doi [\aap] {10.1051/0004-6361/202243472}, \href
  {https://ui.adsabs.harvard.edu/abs/2022A&A...664A.157S} {664, A157}

\bibitem[\protect\citeauthoryear{{Tang}, {MacFadyen}  \& {Haiman}}{{Tang}
  et~al.}{2017}]{Yike+2017}
{Tang} Y.,  {MacFadyen} A.,   {Haiman} Z.,  2017, \mn@doi [\mnras]
  {10.1093/mnras/stx1130}, \href
  {https://ui.adsabs.harvard.edu/abs/2017MNRAS.469.4258T} {469, 4258}

\bibitem[\protect\citeauthoryear{{Tang}, {Haiman}  \& {MacFadyen}}{{Tang}
  et~al.}{2018}]{Yike:2018}
{Tang} Y.,  {Haiman} Z.,   {MacFadyen} A.,  2018, \mn@doi [\mnras]
  {10.1093/mnras/sty423}, \href
  {http://adsabs.harvard.edu/abs/2018MNRAS.476.2249T} {476, 2249}

\bibitem[\protect\citeauthoryear{{Tiede}, {Zrake}, {MacFadyen}  \&
  {Haiman}}{{Tiede} et~al.}{2020}]{Tiede:2020}
{Tiede} C.,  {Zrake} J.,  {MacFadyen} A.,   {Haiman} Z.,  2020, \mn@doi [\apj]
  {10.3847/1538-4357/aba432}, \href
  {https://ui.adsabs.harvard.edu/abs/2020ApJ...900...43T} {900, 43}

\bibitem[\protect\citeauthoryear{{Tiede}, {Zrake}, {MacFadyen}  \&
  {Haiman}}{{Tiede} et~al.}{2022}]{Tiede:2022}
{Tiede} C.,  {Zrake} J.,  {MacFadyen} A.,   {Haiman} Z.,  2022, \mn@doi [\apj]
  {10.3847/1538-4357/ac6c2b}, \href
  {https://ui.adsabs.harvard.edu/abs/2022ApJ...932...24T} {932, 24}

\bibitem[\protect\citeauthoryear{{Wang}, {Bai}, {Lai}  \& {Lin}}{{Wang}
  et~al.}{2022}]{WangBaiLai:2022}
{Wang} H.-Y.,  {Bai} X.-N.,  {Lai} D.,   {Lin} D. N.~C.,  2022, \mn@doi [arXiv
  e-prints] {10.48550/arXiv.2212.07416}, \href
  {https://ui.adsabs.harvard.edu/abs/2022arXiv221207416W} {p. arXiv:2212.07416}

\bibitem[\protect\citeauthoryear{{Wang}, {Bai}  \& {Lai}}{{Wang}
  et~al.}{2023}]{WangBaiLai:2023}
{Wang} H.-Y.,  {Bai} X.-N.,   {Lai} D.,  2023, \mn@doi [\apj]
  {10.3847/1538-4357/acac77}, \href
  {https://ui.adsabs.harvard.edu/abs/2023ApJ...943..175W} {943, 175}

\bibitem[\protect\citeauthoryear{{Webbink}}{{Webbink}}{1976}]{Webbink:1976}
{Webbink} R.~F.,  1976, \mn@doi [\apj] {10.1086/154781}, \href
  {https://ui.adsabs.harvard.edu/abs/1976ApJ...209..829W} {209, 829}

\bibitem[\protect\citeauthoryear{{Westernacher-Schneider}, {Zrake}, {MacFadyen}
   \& {Haiman}}{{Westernacher-Schneider}
  et~al.}{2022}]{Westernacher-Schneider:2022}
{Westernacher-Schneider} J.~R.,  {Zrake} J.,  {MacFadyen} A.,   {Haiman} Z.,
  2022, \mn@doi [\prd] {10.1103/PhysRevD.106.103010}, \href
  {https://ui.adsabs.harvard.edu/abs/2022PhRvD.106j3010W} {106, 103010}

\bibitem[\protect\citeauthoryear{{Westernacher-Schneider}, {Zrake}, {MacFadyen}
   \& {Haiman}}{{Westernacher-Schneider}
  et~al.}{2023}]{Westernacher-Schneider:2023}
{Westernacher-Schneider} J.~R.,  {Zrake} J.,  {MacFadyen} A.,   {Haiman} Z.,
  2023, \mn@doi [arXiv e-prints] {10.48550/arXiv.2307.01154}, \href
  {https://ui.adsabs.harvard.edu/abs/2023arXiv230701154W} {p. arXiv:2307.01154}

\bibitem[\protect\citeauthoryear{{Xu} et~al.,}{{Xu}
  et~al.}{2023}]{CPTA-GWB:2023}
{Xu} H.,  et~al., 2023, \mn@doi [Research in Astronomy and Astrophysics]
  {10.1088/1674-4527/acdfa5}, \href
  {https://ui.adsabs.harvard.edu/abs/2023RAA....23g5024X} {23, 075024}

\bibitem[\protect\citeauthoryear{{Zrake}, {Tiede}, {MacFadyen}  \&
  {Haiman}}{{Zrake} et~al.}{2021}]{Zrake+2021}
{Zrake} J.,  {Tiede} C.,  {MacFadyen} A.,   {Haiman} Z.,  2021, \mn@doi [\apjl]
  {10.3847/2041-8213/abdd1c}, \href
  {https://ui.adsabs.harvard.edu/abs/2021ApJ...909L..13Z} {909, L13}

\makeatother
\end{thebibliography}

\appendix
\renewcommand\thefigure{A\arabic{figure}}

\setcounter{figure}{1}
\begin{figure*}
\begin{center}$
\begin{array}{ccc}
\includegraphics[scale=0.35]{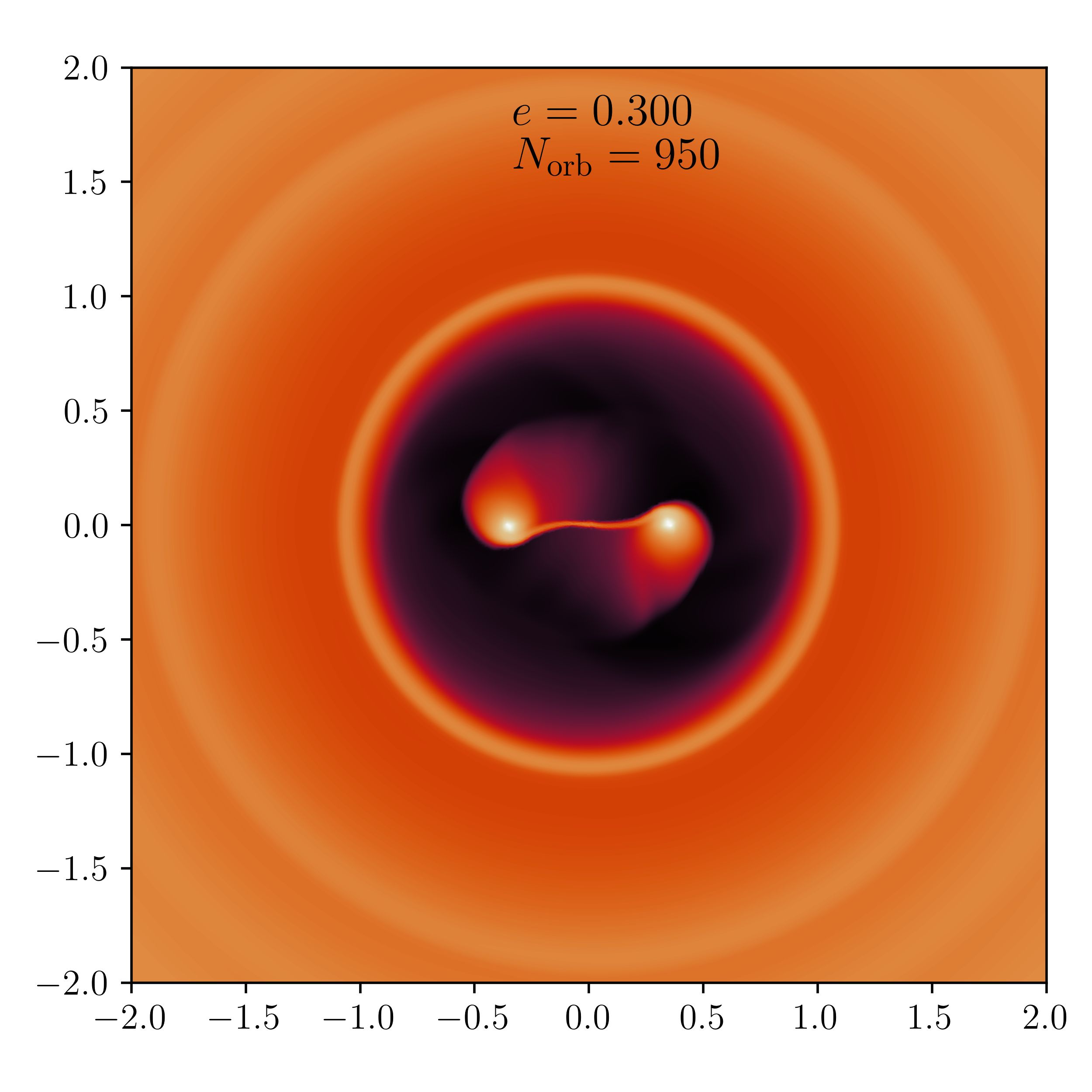} &
 &
 \\
\includegraphics[scale=0.35]{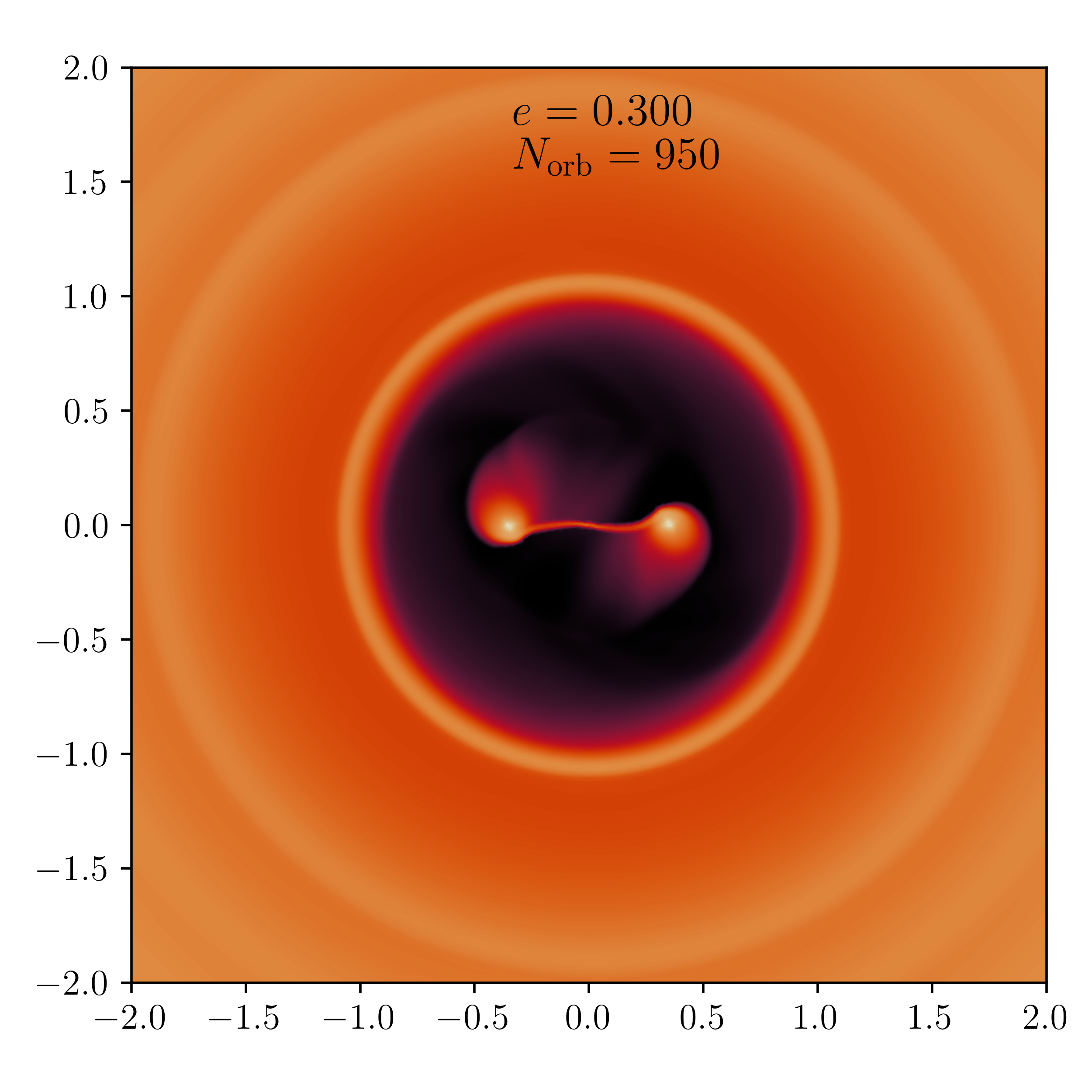} &
\includegraphics[scale=0.35]{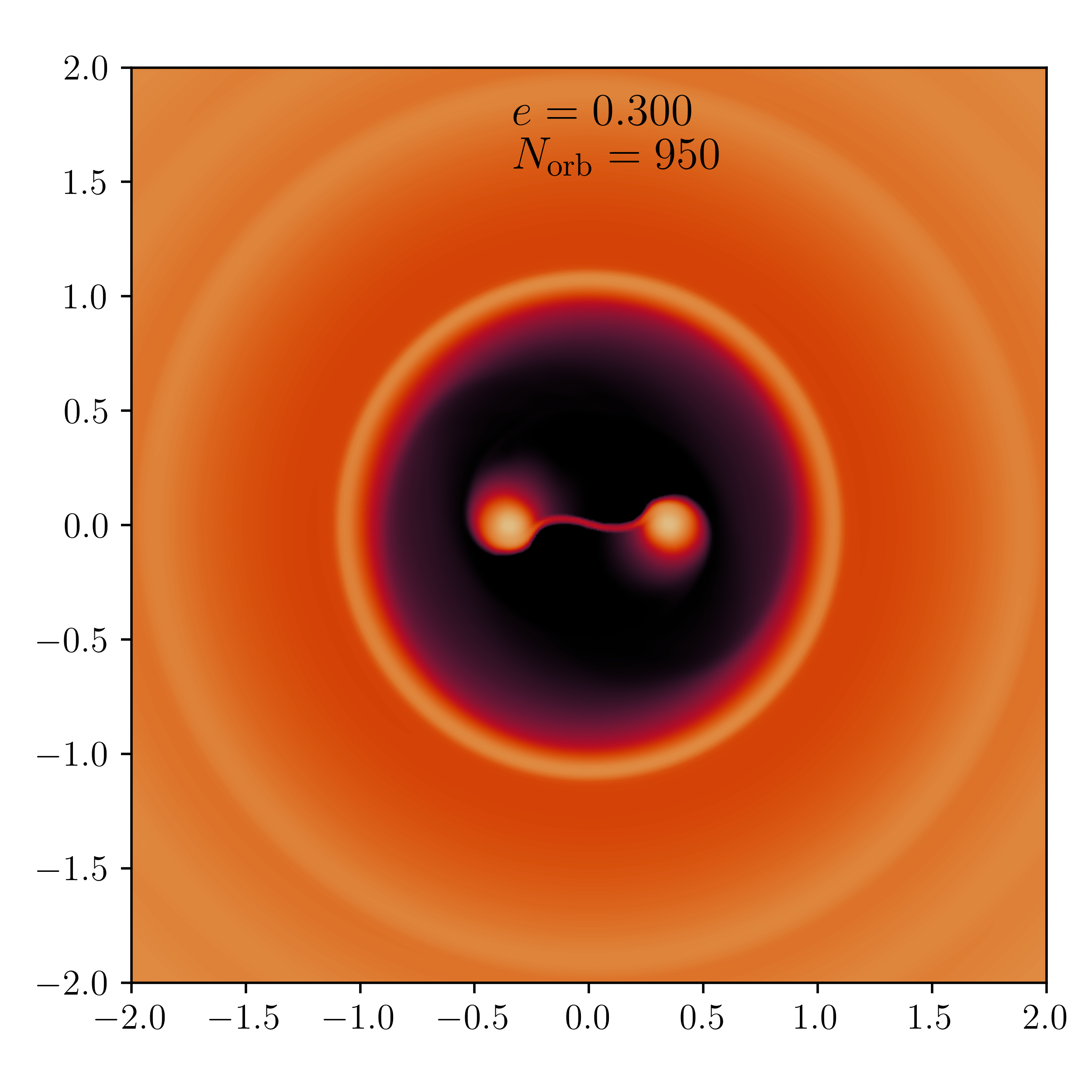} &
\includegraphics[scale=0.35]{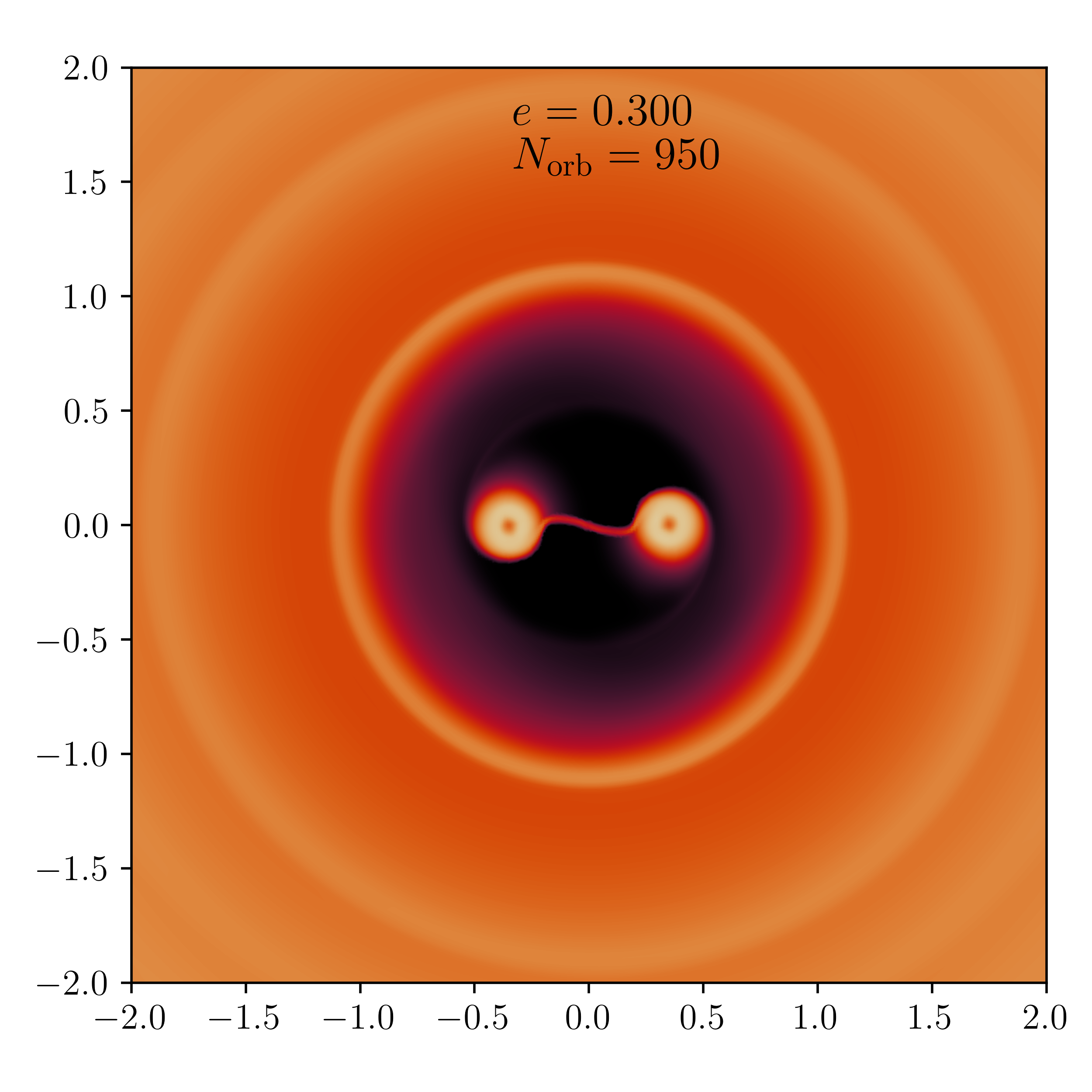} \\
 &
 &
\includegraphics[scale=0.35]{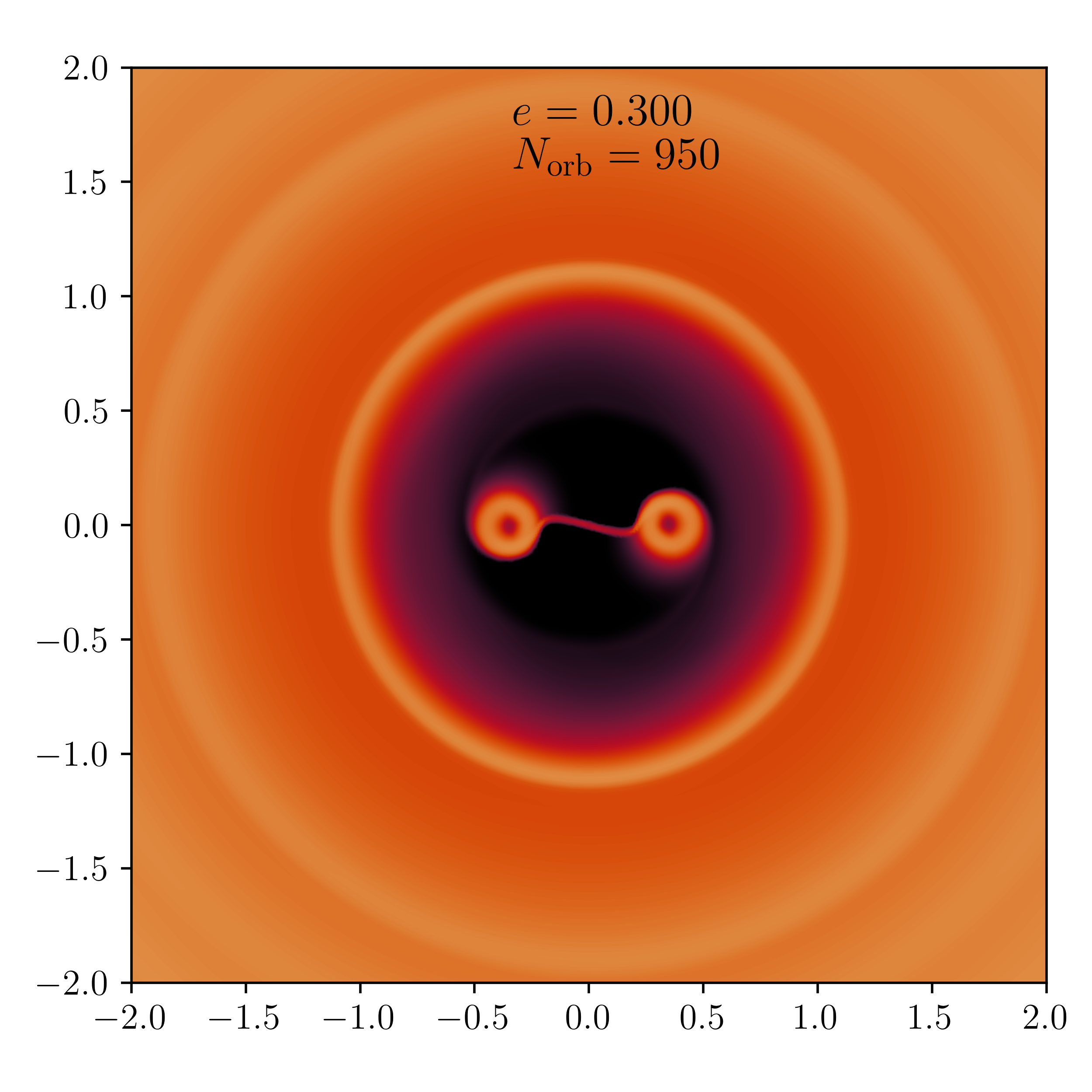} \\
\end{array}$
\end{center}
\vspace{-15pt}
\caption{
Density snapshots for different sink radii and gravitational softening. The diagonal displays density snapshots corresponding to the sink tests in Figure \ref{fig:sink-test}, with $t_s=1\Omega_b$ using the DISCO code (pink dash-dotted lines in \ref{fig:sink-test}). From top left to bottom right, the diagonal changes sink radius and softening together with values ($r_s, r_{\rm soft}$) = ($0.025a$, $0.025a$), ($0.05a$, $0.05a$), ($0.01a$, $0.1a$). The middle column fixes the sink size to $0.05a$, but varies the softening from left to right as $r_{\rm soft} = 0.025a$, $0.05a$, $0.1a$.}
\label{fig:sink_dens}
\end{figure*}

%
\section*{Appendix A: Dependence on  Numerical Parameters} 
\label{A:sinksoft}
%

%
\subsection*{Mass and Momentum Sinks}
\label{a:sink_size_rate}

Unlike in prograde solutions, binaries embedded in a retrograde accretion disk interact directly with the CBD every orbit. Because of this, as noted by \cite{Amaro-Seoane_Retro+2016}, retrograde solutions can be sensitive to the sink prescriptions that remove mass and momentum at each binary component location.
In order to examine the effect of varying the sink implementation on our results, we performed an additional suite of simulations at fixed eccentricity $e=0.3$ in both \texttt{Disco} (\emph{dash-dotted} lines, \emph{circle} markers) and \texttt{Sailfish} (\emph{solid} lines, \emph{cross} markers) where we primarily varied the sink radius $r_{s}$ at two different sink rates $t_s^{-1} = \gamma \Omega_b$ (see definitions in \S \ref{S:Methods}).  
We chose to keep the gravitational softening the same size as the sink radius, however below we relax this assumption as well. 
The results of this study are shown in Appendix Figure \ref{fig:sink-test} where the top-panels show the measured change in binary semi-major axis $\dot a$, the bottom panels show the measured change in eccentricity $\dot e$, the left column shows values derived from both accretion effects and gravitational forces, and the right column shows the effect due only to accreted mass and momentum $\dot a_{\rm acc}$, $\dot e_{\rm acc}$ (note the change in normalization from the measured binary accretion rate to the rate imposed at infinity $\dot M_0$).

The pink lines show the effect of varying $r_s$ at the fiducial sink rate $t_s^{-1} = 1\, \Omega_b$.,
We find that at the fiducial sink-rate our results are largely insensitive to changes in the sink radius, with $\sim 10\%$ change to orbital element rates of changes.
At the fiducial sink rate, there is a small decreasing trend in the magnitude of the total rates with smaller sink radius.
The effects due to accreted mass and momentum $\dot a_{\rm acc}$, $\dot e_{\rm acc}$ do not vary significantly with changing $r_s$.

The cyan lines illustrate the same experiment in the fast-sink regime with $t_s^{-1} = 10\Omega_b$. 
For small enough sink radius $r_s \leq 0.05\,a$, these results for the full $\dot a$ and $\dot e$ appear converged to the fiducial system, and only excessively large sinks $r_s \sim 0.1\,a$ begin to show pathological behavior in the disk morphology and measured orbital evolution.
This can also be seen in the emergence of an outsized contribution to the orbital evolution from the sink terms in the right column of Figure \ref{fig:sink-test}.
We do note, however, that in \texttt{Disco}, while the full results are largely unaffected by the fast sink, 
the disks with fast sinks developed asymmetric and kinked retrograde-bridges, likely leading to the sign change in $\dot e_{\rm acc}$.
\texttt{Disco} runs with fast sinks and $r_s = 0.1\,a$ also developed numerical instabilities that prevented the completion of those runs.

For completeness, we also include one \texttt{Sailfish} run with a ``standard'' acceleration-free sink where $\mathbf{S_J} = S_\Sigma \mathbf{v}$ marked with a black diamond. 
Changing from spinless to acceleration-free sinks similarly does not meaningfully alter our results, and only serves to change the sign of the eccentricity evolution component from the accretion terms alone.

%
\subsection*{Gravitational Softening}

Because the changes in $\dot a_{\rm acc}$, $\dot e_{\rm acc}$ are much smaller than the changes to the total rates (except for the fastest and largest sinks), the changes to the total rates must be primarily from the changing gravitational softening length. To test this, we run a smaller and a larger softening length for the fiducial sink, sampling $r_{\rm soft} = r_s / 2$, $r_s$, $2 r_s$. These are denoted by the triangles in Figure \ref{fig:sink-test}. The primary result, as expected, is that $\dot a_{\rm acc}$, $\dot e_{\rm acc}$ are very weakly, or not affected, while the total rates (left panel) vary on a level similar to that from changing the sink radius above. This is due to the gravitational softening affecting the density structure in the vicinity of the minidisks, and thus, affecting the primary source of gravitational force on the binary (see \S \ref{s:grav-v-acc}).

Figure \ref{fig:sink_dens} shows log-density snapshots for the $e=0.3$ binary sink and softening test, from the \texttt{Disco} runs at fiducial sink rate. From left to right, the gravitational softening is increased from small to large. From top to bottom the sink radius is increased from small to large. Hence, the diagonal from top left to bottom right shows the results of the sink test denoted by the dot-dashed pink lines in Figure~\ref{fig:sink-test}, where softening and sink radius are increased together. The middle row shows the effect of changing the softening while fixing the sink properties (also plotted in Figure~\ref{fig:sink-test}, see legend). It is clear from the density snapshots that the wake structure is affected by softening, thus resulting in different gravitational forces which are the primary reason for differing values in Figure~\ref{fig:sink-test}. While the general behavior remains unaltered, the magnitude of semi-major axis shrinking and eccentricity driving may be sensitive to the chosen gravitational softening.
We note that the strongest effect appears in $\dot a$ from \texttt{Sailfish}, but the increased non-linear features from a smaller softening radius destabilized this run, and it would require higher resolution to characterize accurately.

Note that we do not expect reliable results for smaller softening radii as the fiducial resolution employed (in the \texttt{Disco} runs) covers the diameter of the core of the softened potential by only four cells, or by approximately 16 cells in area. Higher resolution is needed to probe the effects of gas interacting with stronger regions of the point-mass potential. We leave a more detailed exploration of this effect to future work.


\bsp	
\label{lastpage}
\end{document}